\documentclass[draftcls, onecolumn, journal, letterpaper]{IEEEtran}
\usepackage{color}
\usepackage[latin1]{inputenc}
\usepackage{amsmath}
\usepackage{amsfonts}
\usepackage{amssymb}
\usepackage{epsfig}
\usepackage{units}
\usepackage{nicefrac}
\usepackage[noadjust]{cite}
\usepackage{subfigure}
\usepackage{bm}
\usepackage{verbatim}
\usepackage{pstricks, pst-plot, pst-grad, pstricks-add, pst-node, pstricks-add}

\newcommand{\Prob}[0]{\text{Pr}}
\newcommand{\Expected}[1]{\text{E}\left\{#1\right\}}
\newcommand{\Var}[0]{\mathrm{Var}}
\newcommand{\Cov}[0]{\mathrm{Cov}}

\newcommand{\etal}[0]{\textit{et al.}~}
\newcommand{\Set}[1]{\mathcal{#1}}
\newcommand{\StronglyTypicalSet}[0]{\Set{A}_\epsilon^{*(n)}}

\newcommand{\RV}[1]{\mathrm{#1}}
\newcommand{\Vector}[1]{#1}
\newcommand{\Markov}[0]{\leftrightarrow}

\newcommand{\I}[0]{\mathrm{I}}
\newcommand{\h}[0]{\mathrm{h}}
\newcommand{\C}[0]{\mathrm{C}}
\newcommand{\inverseOrder}[0]{\phi}
\newcommand{\drawniid}[0]{\stackrel{n}{\sim}}

\newcommand{\Vfactor}[0]{\alpha}
\newcommand{\Wfactor}[0]{\beta}
\newcommand{\Vpower}[0]{\Gamma}
\newcommand{\VpowerCov}[0]{{\tilde \Gamma}}
\newcommand{\Wpower}[0]{\Lambda}
\newcommand{\Matrix}[1]{\mathrm{\mathbf{#1}}}
\newcommand{\MatrixElement}[3]{\left[#3\right]_{#1, #2}}
\newcommand{\PathlossExponent}{\theta}

\newcommand{\ShowFigure}[1]{#1}

\newtheorem{lemma}{Lemma}
\newtheorem{theorem}{Theorem}
\newtheorem{remark}{Remark}

\newtheorem{definition}{Definition}
\newtheorem{corollary}{Corollary}

\usepackage[english]{babel}

\title{Analysis of a Mixed Strategy for Multiple Relay Networks}
\author{
  Peter~Rost,~\IEEEmembership{Student~Member,~IEEE,}~and~Gerhard~Fettweis,~\IEEEmembership{Senior~Member,~IEEE}%
  \thanks{Manuscript submitted \today.}%
  }
\pubid{$Id: ieee.transit.tex,v 1.5 2007/09/19 12:19:45 rost Exp $}
\begin{document}
\maketitle
  \begin{abstract}
    In their landmark paper Cover and El Gamal proposed different coding strategies for the
    relay channel with a single relay supporting a communication pair. 
    These strategies are the \textit{decode-and-forward} and \textit{compress-and-forward} approach, as well as a general lower bound 
    on the capacity of a relay network which relies on the mixed application of the previous two strategies. 
    So far, only parts of their work - the decode-and-forward
    and the compress-and-forward strategy - have been applied to networks with multiple relays.

    This paper derives a mixed strategy for multiple relay networks using
    a combined approach of partial decode-and-forward with $N+1$ levels and the ideas of
    successive refinement with different side information at the receivers. After describing
    the protocol structure, we present the achievable rates for the discrete memoryless relay channel
    as well as Gaussian multiple relay networks. Using these results we compare the mixed strategy
    with some special cases, e.\,g., multilevel decode-and-forward, distributed compress-and-forward
    and a mixed approach where one relay node operates in decode-and-forward and the other
    in compress-and-forward mode.
  \end{abstract}
  \section{Introduction}
    The growing popularity of mobile communications systems, sensor networks as well as ad hoc networks draws
    an increased activity in the field of network information theory and relaying
    in particular.  The basic idea of relaying  \cite{vdMeulen.TR.1968,Meulen.AAP.1971}
    is to support a communication pair using additional radios.
    Two basic coding strategies for the one-relay case were proposed by
    Cover and El Gamal in \cite{Cover.Gamal.TransIT.1979}:
    \emph{decode-and-forward} (DF) and \emph{compress-and-forward} (CF); both still serve as 
    basic building blocks of recent protocols. 
    
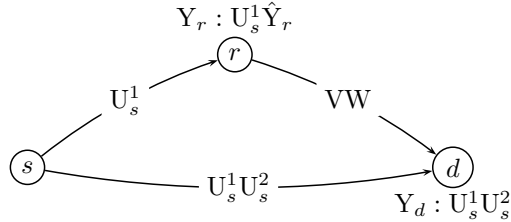
\begin{figure}
      \centering
      \ShowFigure{%
	\begingroup
\unitlength=1mm
\begin{picture}(68, 28.5)(0, 3.5)

  \psset{xunit=1mm, yunit=1mm, linewidth=0.2mm}

  \rput(2, 10){\cnodeput(0, 0){Source}{$s$}}

  \rput(29, 25){
    \cnodeput(0, 0){R1}{$r$}
    \rput(0, 5){$\RV{Y}_r : \RV{U}_s^1\hat{\RV{Y}}_r$}}

  \rput(58, 10){
    \cnodeput(0, 0){Destination}{$d$}
    \rput(0, -5){$\RV{Y}_d:\RV{U}_s^1\RV{U}_s^2$}}
  \ncarc[arcangle=350]{->}{Source}{Destination}\mput*{$\RV{U}_s^1\RV{U}_s^2$}
  \ncarc[arcangle=10]{->}{R1}{Destination}\mput*{$\RV{V}\RV{W}$}
  \ncarc[arcangle=10]{->}{Source}{R1}\mput*{$\RV{U}_s^1$}
\end{picture}
\endgroup
	\caption{Information flow of the general lower bound for the one-relay case.}
	\label{fig:intro:pdf}
      }
    \end{figure}

    Furthermore, [3, Theorem 7] provides a general lower bound on the
    capacity for the one-relay case which can be achieved using a combination of DF and CF.
    More specifically, the source message is divided into two parts: $\RV{U}_s^1$, which is decoded by the relay node, and $\RV{U}_s^2$, 
    which can only be decoded if $\RV{U}_s^1$ is known.  As illustrated in Fig. \ref{fig:intro:pdf}, 
    the relay terminal $r$ decodes $\RV{U}_s^1$ and selects a message $\RV{V}$ using the random 
    binning procedure introduced in \cite{Slepian.Wolf.1973}. Since the source node knows this mapping it
    also knows the message $\RV{V}$ and can support the relay transmitting $\RV{V}$. The relay further uses $\hat{\RV{Y}}_r$ to
    quantize its uncertainty about $\RV{U}_s^2$ in its channel output $\RV{Y}_r$ and selects
    a message $\RV{W}$ depending on this quantization. 
    The destination node $d$ decodes $\RV{V}$ and uses this additional information to decode
    $\RV{U}_s^1$. Using the message $\RV{W}$ and the correlation between the
    relay and destination channel output, the destination decodes the quantization $\hat{\RV{Y}}_r$
    (a strategy similar to Wyner-Ziv coding \cite{Wyner.Ziv.1976}).
    With the quantized channel output of the relay and its own channel output,
    the destination finally decodes the second source message $\RV{U}_s^2$. 

    \subsection{Motivation}
      More recent work on relaying concentrated on the analysis of networks with multiple relays:
      among others, Gupta and Kumar derived in \cite{Gupta.Kumar.TransIT.2000} bounds
      on the capacity of ad-hoc networks without cooperation between individual nodes.
      Again, Gupta and Kumar generalized in \cite{Gupta.Kumar.TransIT.2003} the DF approach to a multilevel relaying scenario 
      where each node decodes the full source message and uses irregular encoding.
      Later, Xie and Kumar proposed in \cite{Xie.Kumar.TransIT.2004,Xie.Kumar.TransIT.2005} a DF strategy based
      on regular encoding and successive decoding, which in general achieves higher rates than the proposal
      in \cite{Gupta.Kumar.TransIT.2003}.
      Besides, Kramer \etal \cite{Kramer.Gastpar.Gupta.TransIT.2005} derived different DF or CF based
      strategies for a variety of different relay networks such as the multiple access and broadcast relay channel.

      Except for \cite{Kramer.Gastpar.Gupta.TransIT.2005}, these approaches regard coding schemes operating
      either in DF or CF mode. In \cite[Theorem 4]{Kramer.Gastpar.Gupta.TransIT.2005} a mixed approach was presented
      where the network consists of both CF and DF nodes. Furthermore, in  
      \cite[Theorem 5]{Kramer.Gastpar.Gupta.TransIT.2005} a specific mixed protocol for two relays was presented. 
      The motivation of this work is the development of a framework generalizing the previously mentioned approaches
      such that the protocol with the best performance or lowest complexity can be found for a specific network setup.
      One way is to derive different protocols and compare them to each other which obviously is not an efficient solution.
      Therefore, we follow in this work a similar approach as \cite[Theorem 7]{Cover.Gamal.TransIT.1979} and derive a strategy
      which is able to specialize to different protocols depending on the actual network setup. This might be of special interest
      for instance in ad-hoc networks or mobile communications systems where it is necessary to use
      low complex protocols. Hence, it is beneficial to support a small number of low complex protocols which provide the best performance in
      specific situations instead of using one high-complex protocol.

    \subsection{Contribution and outline of this work}
      In this work we derive such a mixed strategy which generalizes the previously mentioned approaches.
      The main idea is to divide the source message in $N+1$ partial messages (where $N$ denotes the number of relays 
      in the network). Each relay level $l\in[1;N]$ decodes the first $l$ partial messages and quantizes
      the remaining uncertainty in its own channel output. This quantization is then communicated in different
      successive refinement steps to 
      the next levels $l'>l$, which use this quantization to decode the partial
      source messages $\RV{U}_s^k$ with $l'\geq k>l$.

      In Section \ref{sec:system.model} we define the considered network model as well as used
      notations before we describe in Section \ref{sec:protocol} the protocol in detail and show its commonalities
      with the problem of successive refinement with unstructured side information at the receivers as well as
      the degraded message broadcast channel. We further derive in this section the achievable rates for the
      discrete memoryless relay channel before we apply these results to a wireless model in Section \ref{sec:gaussian.channel}.

  \section{System model, notations and definitions}\label{sec:system.model}
    In the following we will use non-italic uppercase letters
    $\RV{X}$ to denote random variables, non-italic lowercase letters $\RV{x}$ to denote
    events of a random variable and italic letters ($N$ or $n$) are used to denote real or complex-valued scalars.
    Ordered sets are denoted by $\Set{X}$, the cardinality of an ordered set is denoted by $\left\|\Set{X}\right\|$
    and $\left[b; b+k\right]$ is used to denote the ordered set of numbers $\left(b, b+1, \cdots, b+k\right)$.
    With $\RV{X}\drawniid p_{\RV{X} | \RV{Y}}(x | y)$ we denote the random $n$-length
    sequence $\left\{\RV{X}[t]\right\}_{t=1}^n$ whose elements $X[t]\in\Set{X}$ are i.i.d. distributed according to some pdf
    $p_{\RV{X} | \RV{Y}}(x | y)$ (in the following we will drop the subscripts if it clear from the context), i.\,e.,
    $p(\RV{x} | y) = \prod_{t=1}^n p(\RV{x}[t] | y[t])$.
    Let $\RV{X}_k$ be a random variable parameterized using $k$ then 
    $\RV{X}_{\Set{C}}$ denotes the vector of all $\RV{X}_k$ with $k\in\Set{C}$ (this applies
    similarly to sets of events). 
    Matrices are denoted by boldface uppercase letters $\Matrix{K}$ and
    the $i$-th row and $j$-th column of matrix $\Matrix{K}$ is denoted by $\MatrixElement{i}{j}{\Matrix{K}}$.
    
    This paper considers a network of $N+2$ nodes: the source node $s=0$, the set of $N$ relays $t\in\Set{R}:=[1;N]$ and 
    the destination node $d=N+1$.
    The discrete memoryless relay channel is defined by the conditional pdf
    $p\left(\RV{\Vector{y}}_{\Set{R}}, \RV{y}_d | \RV{x}_s, \RV{\Vector{x}}_{\Set{R}}\right)$
    over all possible channel inputs
    $\left(\RV{x}_1, \cdots, \RV{x}_{N}, \RV{x}_{s}\right)\in\Set{X}_1\times\cdots\Set{X}_{N}\times\Set{X}_{s}$ and
    channel outputs $\left(\RV{y}_1, \cdots, \RV{y}_{N}, \RV{y}_{d}\right)\in\Set{Y}_1\times\cdots\Set{Y}_{N}\times\Set{Y}_{d}$
    with $\Set{X}_i$ and $\Set{Y}_j$ denoting the input and output alphabets.

    We will use in the following $\left(\RV{x}^n, \RV{y}^n\right)\in\StronglyTypicalSet\left(\RV{X}, \RV{Y}\right)$ to indicate that the $n$-length sequence tupel
    $\left(\RV{x}^n, \RV{y}^n\right)\in\Set{X}^n\times\Set{Y}^n$ is $\epsilon$-strongly typical with respect to the joint pdf
    $p(\RV{x}, \RV{y})$ where we abbreviate $\StronglyTypicalSet\left(\RV{X}, \RV{Y}\right)$ in the following by $\StronglyTypicalSet$
    if it is clear from the context \cite[Ch. 13.6]{Cover.Thomas.1991}.
    Let $\pi(\Set{X})$ be the set of all permutations of a set $\Set{X}$.
    The source chooses an ordering $o_s\in\pi([1;N+1])$ where
    $o_s(l)$ denotes the $l$-th element of $o_s$ and $o_s(N+1)=N+1$. For the sake of readability
    we abbreviate in the following $\RV{Y}_{o_s(l)}$ by $\RV{Y}_l$ and 
    the relay node $o_s(l)$ by $l$ or as the $l$\emph{-th level}. 
    Besides, each relay $l$ introduces an ordering $o_l\in\pi([l+1; N+1])$ where $o_l(i)$
    indicates node $o_s(o_l(i))$. We further use in the following the
    function $\inverseOrder_l$ to denote the inverse of $o_l$, i.\,e., $o_l(\inverseOrder_l(i))=i$.
    \begin{remark}
      In comparison to \cite{Gupta.Kumar.TransIT.2003} we do not consider any grouping approach
      where multiple relays are operating simultaneously in one group. The qualitative result
      of this work would not change by grouping nodes but only makes the analysis more involved.
    \end{remark}

    \begin{definition}
      A $(2^{nR_s^1}, 2^{nR_s^2}, \dots, 2^{nR_s^{N+1}}, n, \lambda_n)$ code for the previously
      described system model consists of the following
      \begin{itemize}
	\item A set of equally probable indices $\Set{W}=[1; 2^{nR}]$ where $R\leq\sum_{k=1}^{N+1} R_s^k$ and the corresponding
	  r.v. $\RV{W}$ over $\Set{W}$.
	\item The source encoding function $f_0: [1; 2^{nR}]\rightarrow \Set{X}_s^n$.
	\item The relay encoding functions $f_{l; b}: \Set{Y}_l^{[1; b-1]}\rightarrow \Set{X}_l^n$, $l\in[1;N], b\in[1;B]$, such that
	  the $n$-length sequence $x_l(b)$ in block $b$ is given $x_l(b)=f_{l;b}(y_l(1), \cdots, y_l(b-1))$.
	\item The decoding function $g: \Set{Y}_d^n\rightarrow [1; 2^{nR}]$.
	\item The maximum probability of error 
	  \begin{equation}
	    \lambda_n = \max\limits_{w\in\Set{W}}\Prob\left\{g(\RV{y}_d)\neq w | \RV{W}=w\right\}.
	  \end{equation}
      \end{itemize}
    \end{definition}

    \begin{definition}
      A rate $R$ is said to be achievable if there exists a sequence of
      $(2^{nR_s^1}, \dots, 2^{nR_s^{N+1}}, n, \lambda_n)$ codes with $\sum_{k=1}^{N+1} R_s^k\geq R$
      such that $\lambda_n\rightarrow0$ as $n\rightarrow\infty$.
    \end{definition}

  \section{Protocol description and achievable rates}\label{sec:protocol}
    The encoding procedure of the mixed strategy is based on a block Markov superposition coding in $B$ blocks and utilizes
    the following messages:
    \begin{itemize}
      \item the source messages $\RV{U}_s^k$, $k\in[1;N+1]$, with rates $R_s^k$,
      \item the support messages $\RV{V}_l^k$, $l\in[1;N], k\in[1;l]$, sent by relay level $l$ with rates $R_s^k$ to support the source message $\RV{U}_s^k$,
      \item the quantizations $\RV{\hat Y}_l^{k'}$, $k'\in[1;M_l]$ with $M_l=N-l+1$, used to quantize the channel output $\RV{Y}_l$ in $M_l$
	successive refinement steps, and
      \item the broadcast messages $\RV{W}_l^{k'}$ with rates $\hat R_l^{k'}$, $k'\in[1;M_l]$, to communicate the quantizations.
    \end{itemize}
    The actual channel input $\RV{X}_l$ at node $l$ is then a deterministic function of the previously mentioned messages.
    We use further the regular encoding approach presented in \cite{Xie.Kumar.TransIT.2005}
    as it achieves in general higher rates than the irregular approach proposed in \cite{Gupta.Kumar.TransIT.2003}.
    Consider the following description which explains the encoding and decoding procedure in some arbitrary block $b\in[1;B]$
    and is illustrated in Fig. \ref{fig:protocol:information.exchange} for $N=2$:
    \begin{figure}
      \centering
      \ShowFigure{%
	\begingroup
\unitlength=1mm
\begin{picture}(88, 55)(0, 0)

  \psset{xunit=1mm, yunit=1mm, linewidth=0.2mm}

  \rput(2, 10){\cnodeput(0, 0){Source}{$s$}}

  \rput(25, 45){
    \cnodeput(0, 0){R1}{$1$}
    \rput(0, 7){$\RV{Y}_1 : \RV{U}_s^1\hat{\RV{Y}}_1^1\hat{\RV{Y}}_1^2$}}
  \ncarc{->}{Source}{R1}\mput*{$\RV{U}_s^1$}

  \rput(65, 45){
    \cnodeput(0, 0){R2}{$2$}
    \rput(0, 7){$\RV{Y}_2 : \RV{U}_s^1\RV{U}_s^2\hat{\RV{Y}}_2^1$}}
  \ncarc{->}{R1}{R2}\mput*{$\RV{V}_1^1\RV{W}_1^1\RV{W}_1^2$}
  \ncarc{->}{Source}{R2}\mput*{$\RV{U}_s^1\RV{U}_s^2$}

  \rput(77, 10){\cnodeput(0, 0){Destination}{$d$}\rput(0, -6){$\RV{Y}_d:\RV{U}_s^1\RV{U}_s^2\RV{U}_s^3$}}
  \ncarc{->}{Source}{Destination}\mput*{$\RV{U}_s^1\RV{U}_s^2\RV{U}_s^3$}
  \ncarc{->}{R1}{Destination}\mput*{$\RV{V}_1^1\RV{W}_1^1$}
  \ncarc{->}{R2}{Destination}\mput*{$\RV{V}_2^1\RV{V}_2^2\RV{W}_2^1$}
\end{picture}
\endgroup
	\caption{\emph{Information exchange} of the mixed strategy for $N=2$.}
	\label{fig:protocol:information.exchange}
      }
    \end{figure}
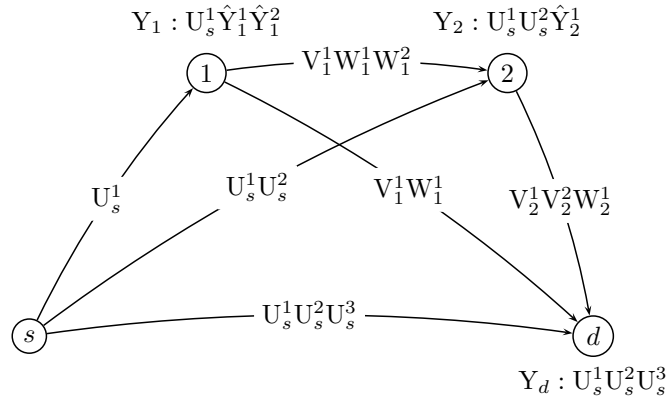
    \paragraph{\textbf{Encoding in block $\bm{b}$}}
      The source divides its message into $N+1$ partial messages $\RV{U}_s^k$. Furthermore,
      as the source node knows the mapping of source to relay messages applied at each relay node,
      it can support the relay messages, e.\,g., using a coherent transmission. 

      In block $b$ the relay level $l$ supports the source messages sent in block $b-l$. Hence,
      assume that it decoded all source messages $\RV{U}_s^{[1;l]}$ sent in block $b-l$.
      Then, it can use the decoded indices to select the corresponding messages from its own codebook.
      We assume that the relay also correctly decoded all source messages sent in blocks $[b-N; b-l-1]$. 
      Therefore, relay $l$ knows all messages $\RV{V}_{l'}^{[1;l]}$ sent by levels $l'>l$ (as all codebooks were
      revealed to each relay node) and can support these subsequent levels.
      Furthermore, the relay node compresses the remaining uncertainty in its channel output by the quantizations
      $\RV{\hat Y}_l^{[1;M_l]}$ where node $o_l(k)$ decodes all $\RV{\hat Y}_l^{[1;k]}$ using the
      broadcast messages $\RV{W}_l^{[1;k]}$ sent by level $l$.

    \paragraph{\textbf{Decoding in block $\bm{b}$}}
      Consider level $l\in[1;N+1]$ and the decoding procedure of the partial source message $\RV{U}_s^k$ sent in block $b-l+1$. 
      Note that level $l$ decodes the partial messages in ascending order such that when decoding $\RV{U}_s^k$ all messages
      used to decode $\RV{U}_s^{[1;k-1]}$ are known. 
      At first, level $l$ decodes the broadcast messages $\RV{W}_{k-1}^{[1;\inverseOrder_{k-1}(l)]}$ sent by level $k-1$ in block $b-l+k$.
      Using these messages the relay can decode the quantization $\RV{\hat Y}_{k-1}^{\inverseOrder_{k-1}(l)}$ for block $b-l+1$
      using its own channel output $\RV{Y}_l$ as side information. 
      
      The decoding of the source messages $\RV{U}_s^k$ is done by searching for
      an unique message index such that the source message sent in block $b-l+1$ and all support messages $\RV{V}_{l-m}^k$,
      $m\in[1;l-k]$, sent in block $b-m+1$ are jointly typical with the quantizations $\RV{\hat Y}_{l'\in[1;k-1]}^{\inverseOrder_{l'}(l)}$
      and the own channel output $\RV{Y}_l$ in block $b-l+1$.
      \begin{remark}
	Before we can proceed we must mention that there two substantially different ways of communicating
	quantization messages. The first alternative is to communicate quantizations for block $b-l+1$ after decoding
	all source messages $q_{s,b-l+1}^{[1;l]}$. These quantizations can be used by all subsequent levels to
	decrease the set of jointly $\epsilon$-typical source messages sent in block $b-l+1$. An alternative is to
	communicate in block $b$ the quantizations for the channel output $\RV{y}_l(b-1)$. This has the advantage
	that these quantizations can be used to decrease the set of jointly $\epsilon$-typical relay messages. Furthermore,
	using the latter one we only know the messages sent by level $l-1$, hence the quantization is contains significant
	interference. We decided to use the former alternative as it is easier to generalize and seems to
	offer more benefits (beside the fact that it is less complex).
      \end{remark}

\begin{figure}
	\centering
	\scalebox{0.95}[0.95]{\begingroup
\unitlength=1mm
\begin{picture}(83, 55)(-8, 0)

  \psset{xunit=1mm, yunit=1mm, linewidth=0.2mm}

  \rput(12, 9){\rnode{FL}{\psframebox[framearc=.3]{$f_l^{M_l}$}}}
  \rput(62.05, 10){\rnode{GL}{\psframebox[framearc=.3]{$g_{o_l(M_l)}$}}}
  \rput(62.05, 2){\rnode{YL}{$\RV{Y}_{o_l(M_l)}$}} \ncline{->}{YL}{GL}
  \rput[l](72, 10){\rnode{XL}{$\left(\RV{\hat Y}_l^1, \cdots, \hat{\RV{Y}}_l^{M_l}\right)$}} \ncline{->}{GL}{XL}
  \rput(62.05, 20){$\vdots$} \rput(12, 20){$\vdots$}
  \pnode(55.5, 11){GL1} \pnode(55.5, 10){GL2} \pnode(55.5, 9){GL3}

  \rput(12, 35){\rnode{F2}{\psframebox[framearc=.3]{$f_l^2$}}}
  \rput(62.05, 35){\rnode{G2}{\psframebox[framearc=.3]{$g_{o_l(2)}$}}}
  \rput(62.05, 27){\rnode{Y2}{$\RV{Y}_{o_l(2)}$}} \ncline{->}{Y2}{G2}
  \rput[l](72, 35){\rnode{X2}{$\left(\RV{\hat Y}_l^1, \hat{\RV{Y}}_l^2\right)$}} \ncline{->}{G2}{X2}
  \pnode(56.3, 36){G21}

  \rput(12, 50){\rnode{F1}{\psframebox[framearc=.3]{$f_l^1$}}}
  \rput(62.05, 50){\rnode{G1}{\psframebox[framearc=.3]{$g_{o_l(1)}$}}}
  \rput[l](72, 50){\rnode{X1}{$\hat{\RV{Y}}_l^1$}} \ncline{->}{G1}{X1}
  \rput(62.05, 42){\rnode{Y1}{$\RV{Y}_{o_l(1)}$}} \ncline{->}{Y1}{G1}

  \rput(-6, 50){\rnode{Y}{$\RV{Y}_l$}}
  \dotnode(2, 50){XDot1}
  \dotnode(2, 35){XDot2}
  \ncline{->}{Y}{F1}
  \ncline{-}{XDot1}{XDot2}
  \ncline{->}{XDot2}{F2}
  \ncangle[angleA=90, angleB=180]{->}{XDot2}{FL}

  \dotnode(45, 50){F1Dot}\ncline{->}{F1}{G1}\aput(0.3){$\hat R_l^1$}
  \dotnode(45, 36){F2Dot}\ncline{-}{F1Dot}{F2Dot}
  \ncline{->}{F2Dot}{G21}
  \ncangle[angleA=90, angleB=180]{->}{F2Dot}{GL1}

  \dotnode(48, 35){F2Dot}\ncline{->}{F2}{G2}\aput(0.3){$\hat R_l^2$}
  \ncangle[angleA=90, angleB=180]{->}{F2Dot}{GL2}

  \ncline{->}{FL}{GL3}\aput(0.3){$\hat R_l^{M_l}$}
\end{picture}
\endgroup}
	\caption{The successive refinement problem as it emerges in our protocol. $\RV{Y}_l$
	needs to be quantized by encoders $f_{l}^1, \cdots, f_l^{M_l}$ at different distortions 
	and to be decoded by decoders $g_{o_l(1)}, \cdots, g_{o_l(M_l)}$, $M_l=N-l+1$, which can exploit
	their own channel output as side information.
	}
	\label{figure:protocol:multiple.description.problem}
      \end{figure}
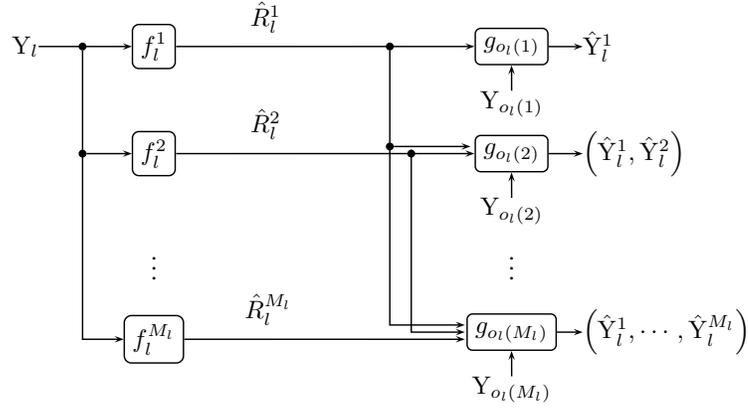

    \paragraph{\textbf{The successive refinement problem}}\label{sec:protocol.structure:multiple.description}
      The previous description already reveals that the quantization of the relay channel output can be described
      as a successive refinement problem \cite{Koshelev.1980,Equitz.Cover.TransIT.1991} with unstructured side information at the receivers.
      Fig. \ref{figure:protocol:multiple.description.problem} illustrates the successive
      refinement problem as it emerges in our proposal. The channel output of level $l$, i.\,e., $\RV{Y}_l$,
      has to be encoded and transmitted to nodes $o_l(i)$, $i\in[1; M_l]$.
      At first the channel output is estimated by a quantization $\hat{\RV{Y}}_l^1$ with rate $\Delta_l^1=R_{\RV{Y}_l}(D_l^1)$,
      where $R_{\RV{Y}_l}(D)$ is the rate-distortion function for some given distortion $D$.
      This estimation needs to be decoded by all nodes $o_l(i)$.
      Since these nodes can exploit (in general unstructured) side information $\RV{Y}_{o_l(i)}$,
      the necessary rate $\hat R_l^1\leq\Delta_l^1$ to describe  $\RV{Y}_l$ at distortion $D_l^1$
      is given by the Wyner-Ziv source coding rate \cite{Wyner.1975}, i.\,e., 
      $\hat R_l^1=\max\limits_{i\in[1;M_l]} R^{\text{WZ}}_{\RV{Y}_l | \RV{Y}_{o_l(i))}}(D_l^1)$, 
      where $R^{\text{WZ}}_{\cdot|\cdot}(\cdot)$ is the Wyner-Ziv rate-distortion function as defined in \cite{Wyner.Ziv.1976}.
      In the next refinement step all levels $o_l(i)$, $i\geq2$, additionally decode the more accurate description $\hat{\RV{Y}}_l^2$ with $D_l^2<D_l^1$. 
      To describe the refined quantization $\hat{\RV{Y}}_l^2$ additional information at rate $\hat R_l^2$ must be provided. Again from
      rate-distortion theory we know that $\hat R_l^2 + \hat R_l^1\geq\max\limits_{i\in[2;M_l]}R^{\text{WZ}}_{\RV{Y}_l | \RV{Y}_{l,i}}(D_l^2)$.
      In \cite{Koshelev.1980,Equitz.Cover.TransIT.1991} the Markovity condition to achieve
      rate-distortion optimal successive refinements is derived. 
      In our setting this implies the Markov chain
      $\RV{Y}_l\Markov \hat{\RV{Y}}_l^{M_l}\Markov \hat{\RV{Y}}_l^{M_l-1}\Markov \cdots \hat{\RV{Y}}_l^1$.
      With this condition we can ensure that $\hat{\RV{Y}}_l^{k'}$ is at least for one node $o_l(i)$, $i\geq k'$,
      rate-distortion optimal, i.\,e., $\sum_{i=1}^{k'}\hat R_l^{i}=\max\limits_{i\in[k';M_l]}R^{\text{WZ}}_{\RV{Y}_l | \RV{Y}_{l,i}}(D_l^{k'})$.
    
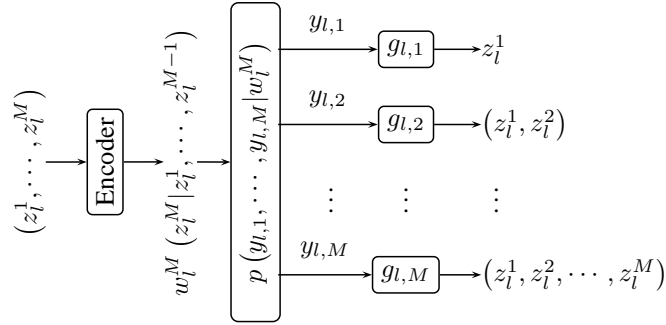
\begin{figure}[t]
	\centering
	\begingroup
\unitlength=1mm
\begin{picture}(87, 40)(3, 10)

  \psset{xunit=1mm, yunit=1mm, linewidth=0.2mm}

  \rput(5, 30){\rnode{Source}{\rotateleft{$\left(z^1_l, \cdots, z^M_l\right)$}}}
  \rput(15, 30){\rnode{Encoder}{\psframebox[framearc=.3]{\rotateleft{Encoder}}}}
  \rput(25, 30){\rnode{Signal}{\rotateleft{$w^M_l\left(z^M_l | z^1_l, \cdots, z^{M-1}_l\right)$}}}
  \ncline{->}{Source}{Encoder}
  \ncline{->}{Encoder}{Signal}

  \rput(35, 30){\rnode{Channel}{\psframebox[framearc=.3]{\rotateleft{$\;\;\;\;p\left(y_{l,1}, \cdots, y_{l,M} | w^M_l\right)\;\;\;\;$}}}}
  \ncline{->}{Signal}{Channel}
  \pnode(38, 45){Channel1}
  \pnode(38, 35){Channel2}
  \pnode(38, 15){ChannelL}

  \rput(55, 45){\rnode{G1}{\psframebox[framearc=.3]{$g_{l,1}$}}}
  \ncline{->}{Channel1}{G1}\naput{$y_{l,1}$}
  \rput[l](65, 45){\rnode{Output1}{$z^1_l$}}
  \ncline{->}{G1}{Output1}
  
  \rput(55, 35){\rnode{G2}{\psframebox[framearc=.3]{$g_{l,2}$}}}
  \ncline{->}{Channel2}{G2}\naput{$y_{l,2}$}
  \rput[l](65, 35){\rnode{Output2}{$\left(z^1_l, z^2_l\right)$}}
  \ncline{->}{G2}{Output2}

  \rput(55, 26){$\vdots$}
  \rput(55, 15){\rnode{GL}{\psframebox[framearc=.3]{$g_{l,M}$}}}
  \rput(45, 26){$\vdots$}
  \ncline{->}{ChannelL}{GL}\naput{$y_{l,M}$}
  \rput(67, 26){$\vdots$}
  \rput[l](65, 15){\rnode{OutputL}{$\left(z^1_l, z^2_l, \cdots, z^M_l\right)$}}
  \ncline{->}{GL}{OutputL}
  
\end{picture}
\endgroup
	\caption{The broadcast channel problem considered in our work. The indices $\Vector{z}_l^{[1;k]}$
	are determined by the Wyner-Ziv coding of the quantizations $\hat{\RV{\Vector{Y}}}_l^{[1;k]}$ and
	need to be decoded by receiver $g_{l,k}, \cdots, g_{l,M_l}$, $M_l=N-l+1$.}
	\label{figure:protocol:broadcast.channel.problem}
      \end{figure}

    \paragraph{\textbf{The broadcast channel problem}}
      Fig. \ref{figure:protocol:broadcast.channel.problem} illustrates the problem of transmitting the quantizations of 
      level $l$, i.\,e., $\hat{\RV{Y}}_l^{[1;M_l]}$, to the next $M_l$ levels: the message indices
      $(z^1_l, \dots, z^{M_l}_l)$, $z^{k'}_l\in[1; 2^{n\hat R_l^{k'}}]$, are determined by the quantizations $\hat{\RV{Y}}_l^1, \dots, \hat{\RV{Y}}_l^{M_l}$ using
      a random binning procedure \cite{Slepian.Wolf.1973}.
      As previously mentioned, nodes $o_l(i)$, $i\in[k'; M_l]$, need the indices $(z^1_l, \dots, z^{k'}_l)$ to successfully decode the quantizations
      $\hat{\RV{Y}}_l^1, \dots, \hat{\RV{Y}}_l^{k'}$.
      Our problem is characterized by a broadcast channel with the degraded message set $\RV{W}^{[1;M_l]}_l$, 
      which was analyzed by Körner and Marton \cite{Koerner.Marton.TransIT.1977}. 
      Using the results of \cite{Koerner.Marton.TransIT.1977} we can state that
      \begin{equation}
	\hat R_l^k \leq \min\limits_{i\in[k;M_l]} \I\left(\RV{W}^k_l; \RV{Y}_{l,i} | \RV{W}^{[1;k-1]}_l\right) \label{eq:protocol:broadcast:100}
      \end{equation}
      is an achievable rate for our problem.
      As explained in \cite[Corollary 5]{Csiszar.Koerner.TransIT.1978},
      (\ref{eq:protocol:broadcast:100}) is included in the capacity region for the case of $M_l=2$:
      $\hat R_l^1 \leq \I(\RV{W}_l^1; \RV{Y}_{l,1})$, $\hat R_l^2 \leq \I(\RV{W}_l^2; \RV{Y}_{l,2} | \RV{W}_l^1)$ and
      $\hat R_l^1 + \hat R_l^2 \leq \I(\RV{W}_l^2; \RV{Y}_{l,2})$ \cite{Koerner.Marton.TransIT.1977}.
      In the special case that $\RV{Y}_{o_l(2)}$ is not ``less noisy'' than $\RV{Y}_{o_l(1)}$ \cite{Koerner.Marton.TransIT.1977},
      i.\,e., $\I(\RV{W}^1_l; \RV{Y}_{l,1})>\I(\RV{W}_l^1; \RV{Y}_{l,2})$,
      we need to introduce a time-sharing 
      and auxiliary random variable to achieve capacity \cite{Csiszar.Koerner.Book.1981}. 
      Since the generalization of this method to prove the capacity region of our setting
      is beyond the scope of this paper we use (\ref{eq:protocol:broadcast:100})
      in the sequel.

      Based on the previous description we can formulate the achievable rates for the mixed strategy in Theorem \ref{theorem:dmc}.
      \begin{theorem}\label{theorem:dmc}
	With the previously presented protocol we are able to achieve any rate 
	\begin{equation}
	  R=\sup\limits_{p}\max\limits_{o_s\in\pi([1;N+1])} \sum\limits_{k=1}^{N+1} R_s^k \label{eq:sdd.mixed:theorem:100}
	\end{equation}
	iff
	\begin{equation}
	  \begin{split}
	    R_s^k < \min\limits_{l\in[k;N+1]}& \I\left(\RV{U}_s^k; \RV{Y}_l, \RV{\hat Y}_{l'\in[1;k-1]}^{\inverseOrder_{l'}(l)} | \RV{U}_s^{[1;k-1]}, 
	    \left\{\RV{V}_{[i;N]}^i, \RV{W}_i^{[1;\inverseOrder_i(l)]}: i\in[1;l]\right\}\right) + \\
	    & \,\sum\limits_{j=k}^{l-1} \I\left(\RV{V}_j^k; \RV{Y}_l | \RV{V}_j^{[1;k-1]},
	    \left\{\RV{V}_i^{[1;i]}, \RV{W}_i^{[1;\inverseOrder_i(l)]} : i\in[j+1;l]\right\}, \RV{V}_{[l;N]}^{[1;l]}\right),
	  \end{split}\label{eq:sdd.mixed:theorem:110}
	\end{equation}
	and $n$ and $B$ are sufficiently large. We further have to constrain the broadcast messages: 
	\begin{align}
	  \begin{split}
	    \hat R_l^j < & \min\limits_{\substack{k\in[j; M_l]\\ k'=o_l(k)}} 
	    \I\left(\RV{W}_l^j; \RV{Y}_{k'} | \RV{V}_l^{[1;l]}, \RV{W}_l^{[1;j-1]}, \left\{\RV{V}_i^{[1;i]}, \RV{W}_i^{[1; \inverseOrder_i(k')]}: i\in[l+1; k']\right\}, \RV{V}_{[k'+1;N]}^{[1;k']}\right),\\
	    & l\in[1;N+1], j\in[1;M_l].\label{eq:sdd.mixed:120}
	  \end{split}
	\end{align}
	Finally, we have the following source coding constraints on the quantizations:
	\begin{align}
	  \begin{split}
	    \sum\limits_{i=1}^m \hat R_l^i > &
	    \max\limits_{\substack{j\in[m;M_l]\\ j'=o_l(j)}}
	      \I\Bigl(\RV{\hat Y}_l^m; \RV{Y}_l | \RV{Y}_{j'}, \RV{\hat Y}_{i\in[1;l-1]}^{\inverseOrder_i(j')},
	      \RV{U}_s^{[1;l]}, \left\{\RV{V}_{[i;N]}^i, \RV{W}_i^{[1;\inverseOrder_i(j')]}: i\in[1;j']\right\}\Bigr),\\
	    & l\in[1; N], m\in[1; M_l].\label{eq:sdd.mixed:130}
	  \end{split}
	\end{align}
	Eq. (\ref{eq:sdd.mixed:120}) and (\ref{eq:sdd.mixed:130}) show the source-channel separation
	in our proposal which is suboptimal in general. The supremum in (\ref{eq:sdd.mixed:theorem:100}) is over all
	pdf of the following form
	\begin{align}
	  \begin{split}
	    & p\left(\RV{u}_s^{[1;N+1]}, \RV{v}_{i\in[1;N]}^{[1;i]}, \RV{w}_{i\in[1;N]}^{[1;M_i]}, \RV{\hat y}_{i\in[1;N]}^{[1;M_i]}, \RV{y}_{[1;N+1]}\right) = \\
	    & p\left(y_{[1;N+1]} | \RV{u}_s^{[1;N+1]}, \RV{w}_{i\in[1;N]}^{[1;M_i]}, \RV{v}_{i\in[1;N]}^{[1;i]}\right)\cdot
	      \prod_{k=1}^{N+1} p\left(\RV{u}_s^k | \RV{u}_s^{[1;k-1]}, \RV{v}_{i\in[k;N]}^k\right)\cdot \\
	    & \prod_{l=1}^N \left[
	      \prod_{k=1}^l p\left(\RV{v}_l^k | \RV{v}_l^{[1;k-1]}, \RV{v}_{[l+1;N]}^k\right)\cdot
	      p\left(\RV{\hat y}_l^{M_l} | \RV{y}_l, \RV{u}_s^{[1;l]}, \RV{v}_{m\in[1;l]}^{[1;m]}, \RV{v}_{[l+1;N]}^{[1;l]}, \RV{w}_{[1;l]}^1\right)
	      p\left(\RV{w}_l^{M_l} | \RV{w}_l^{[1;M_l-1]}, \RV{v}_l^{[1;l]}\right)\cdot\right.\\
	    & \left.\prod_{k=1}^{M_l-1} \left(p\left(\RV{w}_l^k | \RV{w}_l^{[1;k-1]}, \RV{v}_l^{[1;l]}\right)
	      p\left(\RV{\hat y}_l^k | \RV{u}_s^{[1;l]}, \RV{\hat y}_l^{k+1}, \RV{v}_{m\in[1;l]}^{[1;m]}, \RV{v}_{[l+1;N]}^{[1;l]}, \RV{w}_{[1;l]}^1\right)\right)\right].
	  \end{split}\label{eq:sdd.mixed:132}
	\end{align}
      \end{theorem}
      \begin{proof}
	See Appendix \ref{sec:proof:dmc}.
      \end{proof}

  \section{Results for the Gaussian channel}\label{sec:gaussian.channel}
    In the following we apply the previously described mixed strategy to Gaussian multiple relay channels and
    evaluate performance and complexity of different protocols generalized by it.
    
    \subsection{Description of model}
      Again, we consider a network of $N+2$ nodes where
      $d_{l', l}$ denotes the distance between nodes $l'$ and $l\neq l'$. Let $\PathlossExponent$ be the pathloss exponent,
      then the gain factor between both nodes is given in a log-distance path loss model by $h_{l', l}= d_{l',l}^{-\theta/2}$.
      Furthermore, let the channel input at node $l$ be given by $\RV{X}_{l}\drawniid\mathcal{CN}\left(0, P_l\right)$, i.\,e.,
      the channel input is a $n$-length sequence of i.i.d. Gaussian r.v.s with zero mean and variance 
      $\sigma_{\RV{X}}^2 = n^{-1}\sum_{t=1}^n \Expected{\left|\RV{X}_l[t]\right|^2}$. Our analysis uses
      Gaussian alphabets which are not necessarily optimal but easier to treat.
      The channel output at node $l$ and time $t$ is given by
      \begin{equation*}
	\RV{Y}_l[t] = \sum\limits_{l'\in[0;N]\setminus l} h_{l', l}\RV{X}_{l'}[t] + Z_l[t],
      \end{equation*}
      where $Z_l\drawniid\mathcal{CN}\left(0, N_l\right)$ is the additive white Gaussian noise.
      We further use the following messages
      \begin{itemize}
	\item the partial source messages $\RV{U}_s^k\drawniid\mathcal{CN}\left(0, 1\right)$, $k\in[1;N+1]$,
	\item the support messages $\RV{V}_l^k\drawniid\mathcal{CN}\left(0, 1\right)$, 
	  $l\in[1;N], k\in[1;l]$, 
	\item the quantizations $\RV{\hat Y}_l^{k'}\drawniid\mathcal{CN}\left(0, N_l+\sum_{i=k'}^{M_l} N_l^i\right)$, $k'\in[1;M_l]$ with $M_l=N-l+1$,
	  where $\sum_{i=k'}^{M_l} N_l^i$ models the quantization noise to compress the channel output $\RV{Y}_l$,
	\item and the broadcast messages $\RV{W}_l^{k'}\drawniid\mathcal{CN}\left(0, 1\right)$.
      \end{itemize}
      The combination of the partial source messages and the support of the relay transmissions is done as follows:
      \begin{equation}
	\RV{X}_s = \sqrt{P_s}\sum\limits_{k=1}^{N+1} \left(\sqrt{\Vfactor_{s, s}^k}\RV{U}_s^k + \sum\limits_{l'=k}^N\left(\sqrt{\Vfactor_{s,l'}^k}\RV{V}_{l'}^k\right)\right),
      \end{equation}
      where $P_s$ is the source transmission power and $\alpha_{s, l}^k$ is the fraction of power spent by the source node
      for the support of message level $k$ sent by level $l$.
      The channel input at level $l$ is given by
      \begin{equation}
	\RV{X}_l = \sqrt{P_l}\left(\sum\limits_{k=1}^l \sum\limits_{l'=l}^N \sqrt{\Vfactor_{l,l'}^k}\RV{V}_{l'}^k + \sum\limits_{k=1}^{M_l}\sqrt{\Wfactor_l^k}\RV{W}_l^k\right),
      \end{equation}
      where $\Wfactor_l^k$ defines the fraction of power spent by node $l$ for broadcast message level $k$.
      Moreover, to meet the power constraints on all channel inputs, we must ensure that
      \begin{align}
	\sum\limits_{k=1}^{N+1}\left[\Vfactor_{s,s}^k + \sum\limits_{l'=k}^N \Vfactor_{s,l'}^k\right] \leq 1,\text{ and }
	\sum\limits_{k=1}^l\sum\limits_{l'=l}^N \Vfactor_{l,l'}^k + \sum\limits_{k=1}^{M_l} \Wfactor_l^k \leq 1.
      \end{align}
      In the case that no coherent transmission is possible it follows that $\alpha_{l,l'}^k=0$ for $l\neq l'$.

    \subsection{Achievable rates}
      Before we formulate the achievable rates, we need to derive some auxiliary variables.
      Due to the possibility of coherent transmission, the overall received power at level $l'$ for message 
      level $k$ sent by level $l$ is given by $\Vpower_{l,l'}^k$ and defined by
      \begin{align}
	\Vpower_{l,l'}^k & = \left(\sum\limits_{j=k}^{l}\left(h_{j,l'}\sqrt{\Vfactor_{j,l}^k P_j}\right) + h_{s,l'}\sqrt{\Vfactor_{s,l}^k P_s}\right)^2,\label{eq:gauss:140}\\
	\Vpower_{s,l'}^k & = h_{s,l'}^2\Vfactor_{s,s}^kP_s,\text{ and }\Vpower_{l\in\Set{T},l'}^{[1;l]} = \sum\limits_{l\in{\Set{T}}} \Vpower_{l,l'}^{[1;l]}.\label{eq:gauss:141}
      \end{align}
      Furthermore, the cross-correlation of message level $k$ sent by level $l$ and received at nodes $m$ and $m'$ is denoted as 
      $\VpowerCov_{l,m,m'}^k$ and given by 
      \begin{align}
	\begin{split}
	  \VpowerCov_{l,m,m'}^k =\: & \left(\sum\limits_{j\in\left\{[k;l],s\right\}} h_{j,m}\sqrt{\alpha_{j,l}^kP_j}\right)
	    \cdot\left(\sum\limits_{j\in\left\{[k;l],s\right\}} h_{j,m'}\sqrt{\alpha_{j,l}^kP_j}\right),
	\end{split}\label{eq:gauss:142}\\
	\VpowerCov_{s,m,m'}^k =\: & h_{s,m}h_{s,m'}\alpha_{s,s}^k P_s,\text{ and }
	\VpowerCov_{l\in\Set{T},m,m'}^{[k;k']} = \sum\limits_{l\in\Set{T}}\VpowerCov_{l,m,m'}^{[k;k']}.\label{eq:gauss:143}
      \end{align}
      The power of broadcast messages received at node $l$ after decoding the broadcast messages of levels $[k; k']$ is given by
      \begin{align}
	\begin{split}
	  \Wpower_l^{k,k'} = & \sum\limits_{m\in[1;N]\setminus[k;k']} h_{m, l}^2\beta_m^{[1;M_m]}P_m
	    + \sum\limits_{m\in[k;k'-1]}h^2_{m,l}\beta_m^{[\inverseOrder_m(k')+1;M_m]}P_m.
	\end{split}\label{eq:gauss:144}
      \end{align}
      Finally, let us define the matrix $\Matrix{K}_{s,l}^{j,j'}$ as the covariance matrix of $\RV{Y}_l$ and all decoded
      quantizations $\RV{\hat Y}_{l'\in[1; j-1]}^{\inverseOrder_{l'}(l)}$ when decoding the partial source message $\RV{U}_s^j$ and 
      knowing $\RV{U}_s^{[1;j']}$. The elements of the matrix are given by
      \begin{align}
	\MatrixElement{1}{1}{\Matrix{K}_{s,l}^{j,j'}}  =\: & 
	  \Vpower_{l'\in[l+1;N], l}^{[l+1;l']} + \Vpower_{s,l}^{[j'+1;N+1]} + \Wpower_l^{1,l} + N_l,\label{eq:gauss:145}\\
	  \MatrixElement{1+i}{1+i}{\Matrix{K}_{s,l}^{j,j'}} =\: & 
	  \Vpower_{l'\in[l+1;N], i}^{[l+1;l']} + \Vpower_{s,i}^{[j'+1; N+1]} + \Wpower_i^{1,l}
	    + N_i + \sum\limits_{i'=\inverseOrder_i(l)}^{M_i} N_i^{i'},\label{eq:gauss:146}\\
	    \MatrixElement{1}{1+i}{\Matrix{K}_{s,l}^{j,j'}} =\: & 
	  \VpowerCov_{l'\in[l+1;N],l,i}^{[l+1;l']} + \VpowerCov_{s,l,i}^{[j'+1;N+1]},\label{eq:gauss:147}\\
	  \MatrixElement{1+i}{1+i'}{\Matrix{K}_{s,l}^{j,j'}} =\: & 
	  \VpowerCov_{l'\in[l+1;N],i,i'}^{[l+1;l']} + \VpowerCov_{s,i,i'}^{[j'+1;N+1]},\label{eq:gauss:148}
      \end{align}
      for $i,i'\in[1,j-1]$ and $i\neq i'$.
      Using the previously given variables and the specific decoding scheme described above we can formulate Theorem \ref{theorem:gauss}
      on the achievable rates in a Gaussian multiple relay network.
      \begin{theorem}\label{theorem:gauss}
	With the previously presented protocol we are able to achieve any rate 
	\begin{equation}
	  R=\max\limits_{o_s\in\pi([1;N+1])}\sum\limits_{k=1}^{N+1} R_s^k
	\end{equation}
	in the Gaussian multiple relay network iff
	\begin{equation}
	  \begin{split}
	    R_s^k < \min\limits_{l\in[k;N+1]} & \log\left(\frac{\det K_{s,l}^{k-1, k-1}}{\det K_{s,l}^{k-1, k}}\right) + \\
	    &\: \sum\limits_{j=k}^{l-1}\C\left(
	    \frac{\Vpower_{j,l}^k}
	    { \Vpower_{l'\in[1;j-1],l}^{[1;l']} + \Vpower_{j,l}^{[k+1; j]} + \Vpower_{l'\in[l+1;N],l}^{[l+1;l']} + \Wpower_{l}^{j+1,l} + \Gamma_{s,l}^{[1;N+1]} + N_{l} }
	    \right).
	  \end{split}
	\end{equation}
	and $n$ and $B$ are sufficiently large. We further have to constrain the broadcast messages using the following
	upper bound
	\begin{align}
	  \begin{split}
	    \hat R_l^j < & \min\limits_{\begin{array}{l}k\in[j; M_l]\\ k'=o_l(k)\end{array}}\C\left(
	      \frac{h_{l,k'}^2\beta_l^jP_l}
	      {\Vpower_{l'\in[1;l-1],k'}^{[1;l']} + \Vpower_{l'\in[k'+1;N],k'}^{[k'+1;l']} + \Wpower_{k'}^{l+1,k'}-h_{l,k'}^2\sum\limits_{i=1}^{j}\beta_l^iP_l + \Gamma_{s,k'}^{[1;N+1]} + N_{k'}}
	      \right),\\
	    & l\in[1;N+1], j\in[1;M_l],
	  \end{split}
	\end{align}
	Finally, we have the following source coding constraints on the quantizations:
	\begin{equation}
	  \sum\limits_{i=m}^{M_l} N_l^i > \max\limits_{\substack{j\in[m;M_l]\\ j'=o_l(j)}}
	    \frac{
	      \Vpower_{l'\in[j'+1;N], l}^{[j'+1;l']} + \Vpower_{s,l}^{[l+1;N+1]} + \Wpower_l^{1,j'} + N_{l} - 
	      \frac{\left(\VpowerCov_{l'\in[j'+1;N], j', l}^{[j'+1;l']} + \VpowerCov_{s,j',l}^{[l+1;N+1]}\right)^2}
	      {\Vpower_{l'\in[j'+1;N], j'}^{[j'+1;l']} + \Vpower_{s,j'}^{[l+1;N+1]} + \Wpower_{j'}^{1,j'} + N_{j'}}
	    }
	    {2^{\sum\limits_{i=1}^m \hat R_l^i} - 1}.\label{eq:gauss:170}
	\end{equation}
      \end{theorem}
      \begin{proof}
	See Appendix \ref{sec:proof:gauss}.
      \end{proof}

    \subsection{Numerical results}
      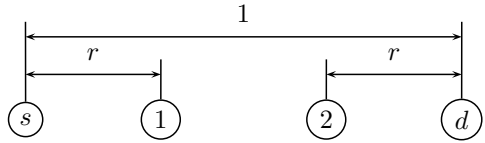
\begin{figure}
	\centering
	\ShowFigure{\begingroup
\unitlength=1mm
\begin{picture}(65, 20)(0, 0)

  \psset{xunit=1mm, yunit=1mm, linewidth=0.2mm}

  \rput(2, 5){\cnodeput(0, 0){Source}{$s$}}
  \rput(60, 5){\cnodeput(0, 0){Destination}{$d$}}
  \rput(20, 5){\cnodeput(0, 0){Relay1}{$1$}}
  \rput(42, 5){\cnodeput(0, 0){Relay2}{$2$}}

  \pnode(2, 18){Source1}
  \pnode(2, 11){Source2}
  \pnode(2, 16){Source3}
  \pnode(60, 18){Destination1}
  \pnode(60, 11){Destination2}
  \pnode(60, 16){Destination3}
  \pnode(20, 13){Relay11}
  \pnode(20, 11){Relay12}
  \pnode(42, 13){Relay21}
  \pnode(42, 11){Relay22}

  \ncline{-}{Source}{Source1}
  \ncline{-}{Destination}{Destination1}
  \ncline{-}{Relay1}{Relay11}
  \ncline{-}{Relay2}{Relay21}

  \ncline{<->}{Source2}{Relay12}\naput{$r$}
  \ncline{<->}{Relay22}{Destination2}\naput{$r$}
  \ncline{<->}{Source3}{Destination3}\naput{$1$}
\end{picture}
\endgroup}
	\caption{Setup for our analysis}
	\label{fig:results:setup}
      \end{figure}
      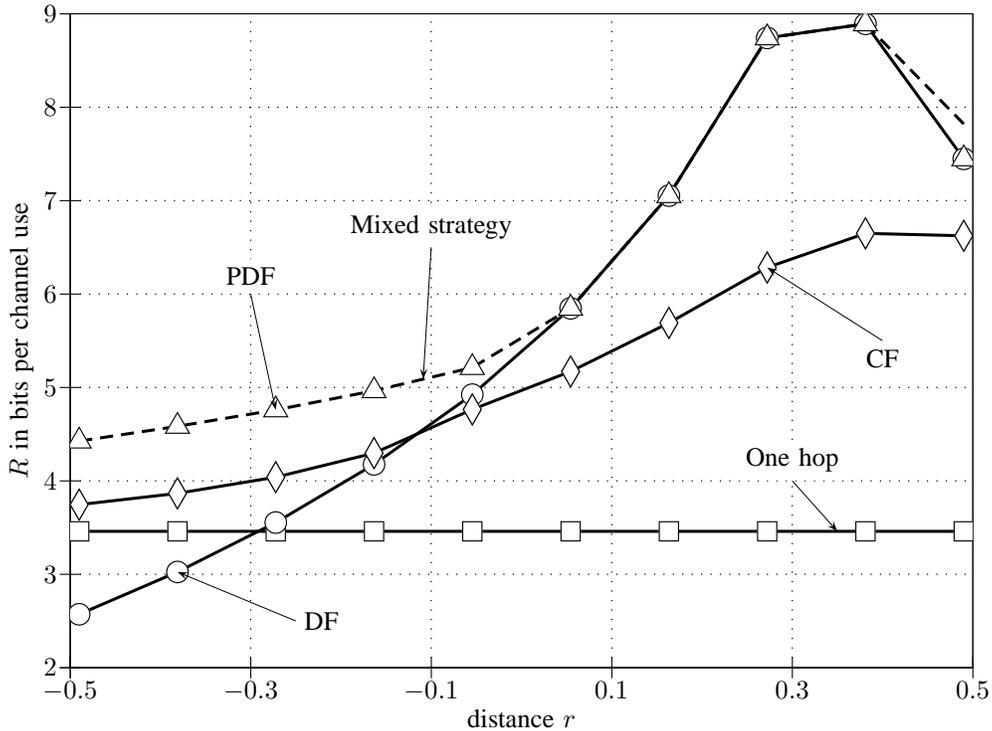
\begin{figure}
	\centering
	\ShowFigure{\begingroup
\unitlength=1mm
\psset{xunit=120mm, yunit=12.4286mm, linewidth=0.1mm}
\psset{arrowsize=2pt 3, arrowlength=1.4, arrowinset=.4}\psset{axesstyle=frame}
\begin{pspicture}(-0.56667, 1.3563)(0.5, 9)
\psaxes[Ox=-0.5, Oy=2, Dx=0.2, Dy=1]{-}(-0.5, 2)(-0.5, 2)(0.5, 9)%
\multips(-0.3, 2)(0.2, 0.0){4}{\psline[linecolor=black, linestyle=dotted, linewidth=0.2mm](0, 0)(0, 7)}
\multips(-0.5, 3)(0, 1){6}{\psline[linecolor=black, linestyle=dotted, linewidth=0.2mm](0, 0)(1, 0)}
\rput[b](0, 1.3563){distance $r$}
\rput[t]{90}(-0.56667, 5.5){$R$ in bits per channel use}
\psclip{\psframe(-0.5, 2)(0.5, 9)}
\psline[linecolor=black, plotstyle=curve, linewidth=0.4mm, showpoints=true, linestyle=solid, dotstyle=square, dotscale=2 2](-0.49, 3.4594)(-0.38111, 3.4594)(-0.27222, 3.4594)(-0.16333, 3.4594)(-0.054444, 3.4594)(0.054444, 3.4594)(0.16333, 3.4594)(0.27222, 3.4594)(0.38111, 3.4594)(0.49, 3.4594)
\psline[linecolor=black, plotstyle=curve, linewidth=0.4mm, showpoints=true, linestyle=solid, dotstyle=o, dotscale=2 2](-0.49, 2.5715)(-0.38111, 3.0244)(-0.27222, 3.5534)(-0.16333, 4.1763)(-0.054444, 4.9227)(0.054444, 5.8465)(0.16333, 7.0529)(0.27222, 8.7415)(0.38111, 8.8918)(0.49, 7.4468)
\psline[linecolor=black, plotstyle=curve, linewidth=0.4mm, showpoints=false, linestyle=dashed](-0.49, 4.4255)(-0.38111, 4.5835)(-0.27222, 4.7594)(-0.16333, 4.9645)(-0.054444, 5.2124)(0.054444, 5.8464)(0.16333, 7.0529)(0.27222, 8.7415)(0.38111, 8.8918)(0.49, 7.8201)
\psline[linecolor=black, plotstyle=curve, linewidth=0.4mm, showpoints=true, linestyle=none, dotstyle=triangle, dotscale=2 2](-0.49, 4.4255)(-0.38111, 4.5835)(-0.27222, 4.7594)(-0.16333, 4.9645)(-0.054444, 5.2124)(0.054444, 5.8465)(0.16333, 7.0529)(0.27222, 8.7415)(0.38111, 8.8918)(0.49, 7.4468)
\psline[linecolor=black, plotstyle=curve, linewidth=0.4mm, showpoints=true, linestyle=solid, dotstyle=diamond, dotscale=2 2](-0.49, 3.7453)(-0.38111, 3.8659)(-0.27222, 4.0393)(-0.16333, 4.2968)(-0.054444, 4.7644)(0.054444, 5.1707)(0.16333, 5.6904)(0.27222, 6.2837)(0.38111, 6.6491)(0.49, 6.623)
\endpsclip
\psline[linecolor=black, linestyle=solid, linewidth=0.1mm]{->}(0.3, 4)(0.35, 3.4594)
\psline[linecolor=black, linestyle=solid, linewidth=0.1mm]{->}(-0.25, 2.5)(-0.38111, 3.0244)
\psline[linecolor=black, linestyle=solid, linewidth=0.1mm]{->}(-0.1, 6.5)(-0.10889, 5.0885)
\psline[linecolor=black, linestyle=solid, linewidth=0.1mm]{->}(-0.3, 6)(-0.27222, 4.7594)
\psline[linecolor=black, linestyle=solid, linewidth=0.1mm]{->}(0.4, 5.5)(0.27222, 6.2837)
\rput[b](0.3, 4){\psframebox[linestyle=none, fillcolor=white, fillstyle=solid]{One hop}}
\rput[l](-0.25, 2.5){\psframebox[linestyle=none, fillcolor=white, fillstyle=solid]{DF}}
\rput[b](-0.1, 6.5){\psframebox[linestyle=none, fillcolor=white, fillstyle=solid]{Mixed strategy}}
\rput[b](-0.3, 6){\psframebox[linestyle=none, fillcolor=white, fillstyle=solid]{PDF}}
\rput[t](0.4, 5.5){\psframebox[linestyle=none, fillcolor=white, fillstyle=solid]{CF}}

\end{pspicture}
\endgroup
 }
	\caption{Achievable rates for \emph{coherent} transmission, $\PathlossExponent=4$, $\gamma_{s,d}=\unit[10]{dB}$ and the setup given in Fig. \ref{fig:results:setup}}
	\label{fig:results:regular_encoding.coherent1.alpha4}
      \end{figure}
      \begin{figure}
	\centering
	\ShowFigure{\begingroup
\unitlength=1mm
\psset{xunit=120mm, yunit=12.4286mm, linewidth=0.1mm}
\psset{arrowsize=2pt 3, arrowlength=1.4, arrowinset=.4}\psset{axesstyle=frame}
\begin{pspicture}(-0.56667, 1.3563)(0.5, 9)
\psaxes[Ox=-0.5, Oy=2, Dx=0.2, Dy=1]{-}(-0.5, 2)(-0.5, 2)(0.5, 9)%
\multips(-0.3, 2)(0.2, 0.0){4}{\psline[linecolor=black, linestyle=dotted, linewidth=0.2mm](0, 0)(0, 7)}
\multips(-0.5, 3)(0, 1){6}{\psline[linecolor=black, linestyle=dotted, linewidth=0.2mm](0, 0)(1, 0)}
\rput[b](0, 1.3563){distance $r$}
\rput[t]{90}(-0.56667, 5.5){$R$ in bits per channel use}
\psclip{\psframe(-0.5, 2)(0.5, 9)}
\psline[linecolor=black, plotstyle=curve, linewidth=0.4mm, showpoints=true, linestyle=solid, dotstyle=square, dotscale=2 2](-0.49, 3.4594)(-0.38111, 3.4594)(-0.27222, 3.4594)(-0.16333, 3.4594)(-0.054444, 3.4594)(0.054444, 3.4594)(0.16333, 3.4594)(0.27222, 3.4594)(0.38111, 3.4594)(0.49, 3.4594)
\psline[linecolor=black, plotstyle=curve, linewidth=0.4mm, showpoints=true, linestyle=solid, dotstyle=o, dotscale=2 2](-0.49, 1.8795)(-0.38111, 2.2586)(-0.27222, 2.7169)(-0.16333, 3.2762)(-0.054444, 3.973)(0.054444, 4.8762)(0.16333, 6.1305)(0.27222, 8.0706)(0.38111, 8.8918)(0.49, 7.4468)
\psline[linecolor=black, plotstyle=curve, linewidth=0.4mm, showpoints=false, linestyle=dashed](-0.49, 3.7499)(-0.38111, 3.8725)(-0.27222, 4.0451)(-0.16333, 4.3385)(-0.054444, 4.7644)(0.054444, 5.2989)(0.16333, 6.311)(0.27222, 8.0706)(0.38111, 8.8918)(0.49, 7.5272)
\psline[linecolor=black, plotstyle=curve, linewidth=0.4mm, showpoints=true, linestyle=solid, dotstyle=triangle, dotscale=2 2](-0.49, 3.7036)(-0.38111, 3.7812)(-0.27222, 3.8892)(-0.16333, 4.0409)(-0.054444, 4.2547)(0.054444, 4.8762)(0.16333, 6.1305)(0.27222, 8.0706)(0.38111, 8.8918)(0.49, 7.4468)
\psline[linecolor=black, plotstyle=curve, linewidth=0.4mm, showpoints=true, linestyle=solid, dotstyle=diamond, dotscale=2 2](-0.49, 3.7453)(-0.38111, 3.8659)(-0.27222, 4.0393)(-0.16333, 4.2968)(-0.054444, 4.7644)(0.054444, 5.1707)(0.16333, 5.6904)(0.27222, 6.2837)(0.38111, 6.6491)(0.49, 6.623)
\endpsclip
\psline[linecolor=black, linestyle=solid, linewidth=0.1mm]{->}(0.3, 4)(0.35, 3.4594)
\psline[linecolor=black, linestyle=solid, linewidth=0.1mm]{->}(-0.1, 2.5)(-0.27222, 2.7169)
\psline[linecolor=black, linestyle=solid, linewidth=0.1mm]{->}(0, 7)(0.10889, 5.8049)
\psline[linecolor=black, linestyle=solid, linewidth=0.1mm]{->}(-0.1, 6)(-0.054444, 4.2547)
\psline[linecolor=black, linestyle=solid, linewidth=0.1mm]{->}(0.4, 5.5)(0.27222, 6.2837)
\rput[b](0.3, 4){\psframebox[linestyle=none, fillcolor=white, fillstyle=solid]{One hop}}
\rput[l](-0.1, 2.5){\psframebox[linestyle=none, fillcolor=white, fillstyle=solid]{DF}}
\rput[b](0, 7){\psframebox[linestyle=none, fillcolor=white, fillstyle=solid]{Mixed strategy}}
\rput[b](-0.1, 6){\psframebox[linestyle=none, fillcolor=white, fillstyle=solid]{PDF}}
\rput[t](0.4, 5.5){\psframebox[linestyle=none, fillcolor=white, fillstyle=solid]{CF}}

\end{pspicture}
\endgroup
 }
	\caption{Achievable rates for \emph{noncoherent} transmission, $\PathlossExponent=4$, $\gamma_{s,d}=\unit[10]{dB}$ and the setup given in Fig. \ref{fig:results:setup}}
	\label{fig:results:regular_encoding.coherent0.alpha4}
      \end{figure}
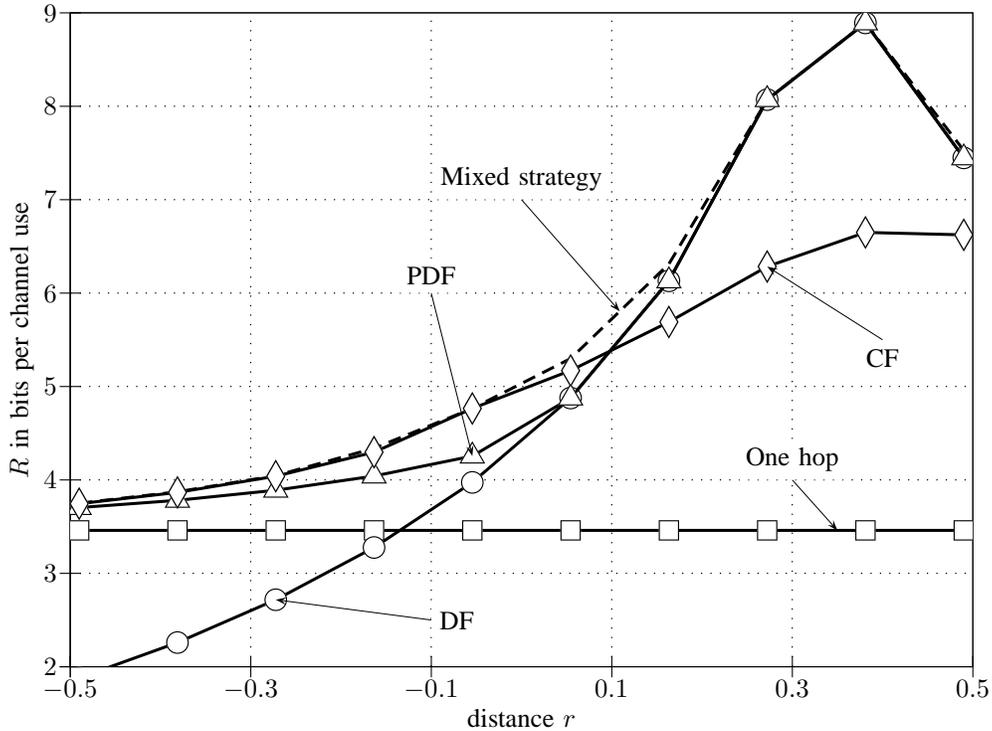
      In this section we apply the previously presented approach to the setup with $N=2$ relay nodes
      illustrated in Fig. \ref{fig:results:setup}. In this setup all distances are normalized to the source
      destination distance, i.\,e., $d_{s,d}=1$, $d_{s,1}=d_{2,d}=\left|r\right|$ and $d_{1,2}=1-2r$. 
      Furthermore, let $N_1=N_2=\cdots N_d$, $P_s=P_1=\cdots P_N$ and 
      $\gamma_{s,d}=\nicefrac{P_s}{N_d}=\unit[10]{dB}$. Other values of $\gamma_{s,d}$
      would impact the quantitative statements given in the following but the qualitative statements remain
      almost unaffected.
      We compare the results of coherent transmission with noncoherent transmission for a pathloss exponent $\PathlossExponent=4$.

      Before we analyze the results for the given setup we need to introduce some special cases of the
      presented approach which, as we see later, achieve its performance in certain scenarios:
      \begin{enumerate}
	\item \textbf{DF}:
	  The pure decode-and-forward case was presented in \cite{Gupta.Kumar.TransIT.2003} where only one source message level
	  is used, i.\,e., only $\RV{U}_s^1$ and $\RV{V}_{l\in[1;N]}^1$ are used. In this case the encoding and decoding complexity
	  of the mixed strategy is minimized.
	\item \textbf{CF}:
	  Assume the source node only uses the partial message $\RV{U}_s^{N+1}$ (by setting $\alpha_{s,s}^{[1;N]}=0$). In this case
	  each relay node only transmits its quantization to the destination node.
	  Hence, the decoding complexity at the destination is increased in comparison to the DF approach. Note that this protocol
	  is comparable to the ideas of \cite[Theorem 3]{Kramer.Gastpar.Gupta.TransIT.2005}.
	\item \textbf{PDF}:
	  This protocol relates to a parameterization with $\beta_{l\in[1;N]}^{[1;M_l]}=0$, i.\,e., the relay nodes
	  do not quantize their channel output. This partial decode-and-forward approach can be seen as a more
	  general form of DF.
	\item \textbf{Mixed CF/DF}:
	  The last special case can be applied for the two-relay case when the first relay node operates in CF mode
	  to support the second relay node which operates in (single-level) DF mode. 
	  In this way the first relay helps to increase the rate
	  towards the second relay which on the other hand can provide more additional information to the destination.
      \end{enumerate}

      Consider Fig. \ref{fig:results:regular_encoding.coherent1.alpha4} which shows the maximum achievable rates
      for direct transmission (one hop), the presented approach and its special cases (except for the mixed CF/DF case which is discussed later)
      in the described setup when coherent transmission is possible. It shows that the PDF approach achieves the performance of the 
      presented strategy at all $r$ except for $r\approx 0.5$.
      This implies that the additional quantization stages provide only minor or no benefits.
      A less complex protocol is DF which for $0\leq r\leq 0.5$
      also achieves the performance of the described strategy at much less encoding and decoding complexity. 
      A closer look on the actual parameters reveals that
      the performance difference between PDF and DF for $r<0$ is due to the fact that in DF all relays must decode the
      source message whereas in PDF an appropriate parameterization can be chosen such that one relay node is not used
      anymore. Hence, for $r<0$ the single hop DF and for $0\leq r\leq 0.5$ the multihop DF
      achieve the best performance while requiring the least encoding and decoding complexity.
      We can further observe in both figures that the CF strategy does not achieve
      the other protocols' performance in the observed interval $-0.5\leq r\leq 0.5$. This coincides with the results for one relay
      where CF can only achieve the general lower bound if the relay is placed closer to the destination (as in this
      region the broadcast cut is the limiting factor). In contrast to the one relay setup, we have in our setup 
      always one relay node with a sufficiently good link to the source node such that CF would only achieve the 
      rate of the mixed strategy for $r\leq-0.5$. 

      Now consider Fig. \ref{fig:results:regular_encoding.coherent0.alpha4} which shows the results when no coherent transmission is possible.
      In this case, the behavior of the presented protocols slightly changes, e.\,g., CF is now able to achieve the performance
      of the mixed strategy for $r\leq$ and both PDF and DF only provide the best performance for $0.2\leq r$. This again
      coincides with the results for the one-relay case where coherent transmission only provides benefits for $r\leq 0.5$.
      Nonetheless, in the more interesting case (from a practical point of view) where the relays are place between source and
      destination, the DF approach again offers the best performance (and performs as good as PDF) which again shows that the
      DF approach offers the best complexity-benefit tradeoff in a wide range of scenarios.

      In both figures $r=0.5$ represents a special point as none of the illustrated protocols achieves the performance
      of our approach. Consider again the mixed CF/DF strategy where the first relay node supports the second relay node
      by communicating its quantized channel output and the second relay node uses these quantizations to operate in single DF mode. At 
      $r=0.5$ this protocol achieves the best performance (the protocol is not shown as it does not provide any advantage
      at other values of $r$). 
    
  \bibliographystyle{IEEEtran}
  \bibliography{IEEEfull,my-references}

\begin{thebibliography}{10}
\providecommand{\url}[1]{#1}
\csname url@rmstyle\endcsname
\providecommand{\newblock}{\relax}
\providecommand{\bibinfo}[2]{#2}
\providecommand\BIBentrySTDinterwordspacing{\spaceskip=0pt\relax}
\providecommand\BIBentryALTinterwordstretchfactor{4}
\providecommand\BIBentryALTinterwordspacing{\spaceskip=\fontdimen2\font plus
\BIBentryALTinterwordstretchfactor\fontdimen3\font minus
  \fontdimen4\font\relax}
\providecommand\BIBforeignlanguage[2]{{%
\expandafter\ifx\csname l@#1\endcsname\relax
\typeout{** WARNING: IEEEtran.bst: No hyphenation pattern has been}%
\typeout{** loaded for the language `#1'. Using the pattern for}%
\typeout{** the default language instead.}%
\else
\language=\csname l@#1\endcsname
\fi
#2}}

\bibitem{vdMeulen.TR.1968}
E.~van~der Meulen, ``Transmission of information in a t-terminal discrete
  memoryless channel,'' Dept. of Statistics, Univ. of California, Berkeley
  (CA), Tech. Rep., 1968.

\bibitem{Meulen.AAP.1971}
------, ``Three-terminal communication channels,'' \emph{Advances in Applied
  Probability}, vol.~3, no.~1, pp. 120--154, 1971.

\bibitem{Cover.Gamal.TransIT.1979}
T.~Cover and A.~E. Gamal, ``Capacity theorems for the relay channel,''
  \emph{{IEEE} Transactions on Information Theory}, vol.~25, no.~5, pp.
  572--584, September 1979.

\bibitem{Slepian.Wolf.1973}
D.~Slepian and J.~Wolf, ``Noiseless coding of correlated information sources,''
  \emph{{IEEE} Transactions on Information Theory}, vol. IT-19, no.~4, pp.
  471--480, July 1973.

\bibitem{Wyner.Ziv.1976}
A.~Wyner and J.~Ziv, ``The rate-distortion function for source coding with side
  information at the decoder,'' \emph{{IEEE} Transactions on Information
  Theory}, vol. IT-22, no.~1, pp. 1--10, January 1976.

\bibitem{Gupta.Kumar.TransIT.2000}
P.~Gupta and P.~Kumar, ``The capacity of wireless networks,'' \emph{{IEEE}
  Transactions on Information Theory}, vol.~46, no.~2, pp. 388--404, March
  2000.

\bibitem{Gupta.Kumar.TransIT.2003}
------, ``Towards an information theory of large networks: An achievable rate
  region,'' \emph{{IEEE} Transactions on Information Theory}, vol.~49, no.~8,
  pp. 1877--1894, August 2003.

\bibitem{Xie.Kumar.TransIT.2004}
L.-L. Xie and P.~Kumar, ``A network information theory for wireless
  communication: Scaling laws and optimal operation,'' \emph{{IEEE}
  Transactions on Information Theory}, vol.~50, no.~5, pp. 748--767, May 2004.

\bibitem{Xie.Kumar.TransIT.2005}
------, ``An achievable rate for the multiple-level relay channel,''
  \emph{{IEEE} Transactions on Information Theory}, vol.~51, no.~4, pp.
  1348--1358, April 2005.

\bibitem{Kramer.Gastpar.Gupta.TransIT.2005}
G.~Kramer, M.~Gastpar, and P.~Gupta, ``Cooperative strategies and capacity
  theorems for relay networks,'' \emph{{IEEE} Transactions on Information
  Theory}, vol.~51, no.~9, pp. 3037--3063, September 2005.

\bibitem{Cover.Thomas.1991}
T.~Cover and J.~Thomas, \emph{Elements of Information Theory}.\hskip 1em plus
  0.5em minus 0.4em\relax John Wiley \& Sons, Inc., 1991.

\bibitem{Koshelev.1980}
V.~Koshelev, ``Hierarchical coding of discrete sources,'' \emph{Problemy
  Peredachi Informatsii}, vol.~16, no.~3, pp. 31--49, 1980.

\bibitem{Equitz.Cover.TransIT.1991}
W.~Equitz and T.~Cover, ``Successive refinement of information,'' \emph{{IEEE}
  Transactions on Information Theory}, vol.~37, no.~2, pp. 269--275, March
  1991.

\bibitem{Wyner.1975}
A.~Wyner, ``On source coding with side information at the decoder,''
  \emph{{IEEE} Transactions on Information Theory}, vol. IT-21, no.~3, pp.
  294--300, May 1975.

\bibitem{Koerner.Marton.TransIT.1977}
J.~Körner and K.~Marton, ``General broadcast channels with degraded message
  sets,'' \emph{{IEEE} Transactions on Information Theory}, vol.~23, no.~1, pp.
  60--64, January 1977.

\bibitem{Csiszar.Koerner.TransIT.1978}
I.~Csiszár and J.~Körner, ``Broadcast channels with confidential messages,''
  \emph{{IEEE} Transactions on Information Theory}, vol.~24, no.~3, pp.
  339--348, May 1978.

\bibitem{Csiszar.Koerner.Book.1981}
I.~Csiszàr and J.~Körner, \emph{Information Theory: coding theorems for
  discrete memoryless channels}, ser. Probability and Mathematical
  Statistics.\hskip 1em plus 0.5em minus 0.4em\relax New York: Academic, 1981.

\end{thebibliography}

  \appendices
  \section{Proof of Theorem \ref{theorem:dmc}}\label{sec:proof:dmc}
  This section proves Theorem \ref{theorem:dmc} using the arguments developed in \cite{Xie.Kumar.TransIT.2005} as well as 
  strongly $\epsilon$-typical sequences as defined in \cite[Ch. 13.6]{Cover.Thomas.1991}. 
  The necessity for strongly $\epsilon$-typical sequences arises from the
  usage of quantizations which requires the Markov lemma \cite[Lemma 14.8.1]{Cover.Thomas.1991}.
  Before we can describe the proof in more detail we need to derive the following lemma which is a basic
  part of this proof.
  \begin{lemma}\label{lemma:sdd:mixed.strategy:2}
    Let the $n$-length tupel $\left(x, y, u, w\right)$ be drawn from
    $\left(\RV{X}, \RV{Y}, \RV{U}, \RV{W}\right)\drawniid p(\RV{x}|\RV{u})p(\RV{y}|\RV{u}, \RV{w})p(\RV{w}, \RV{u})$,
    then
    \begin{equation}
      P_e = \Prob\left\{\left(\RV{x}, \RV{y}, \RV{u}, \RV{w}\right)\in\StronglyTypicalSet\right\}\doteq
      2^{-nI(\RV{X}; \RV{Y}\RV{W}| \RV{U})},
    \end{equation}
    where $\doteq$ is used as defined in \cite{Cover.Thomas.1991}.
  \end{lemma}
  \begin{proof}
    \begin{align}
      P_e & = \sum\limits_{\left(\RV{x}, \RV{y}, \RV{u}, \RV{w}\right)\in\Set{A}_\epsilon^{(n)}}p(\RV{x}|\RV{u})p(\RV{y}|\RV{u}, \RV{w})p(\RV{w}, \RV{u})\\
      & = \left\|\Set{A}_\epsilon^{(n)}\right\| 2^{-n(H(\RV{X} | \RV{U}) - 2\epsilon)}2^{-n(H(\RV{Y} | \RV{W}\RV{U})-2\epsilon)}2^{-n(H(\RV{W}\RV{U}-\epsilon)}\\
      & = 2^{n(H(\RV{X}\RV{Y}\RV{W}\RV{U})+\epsilon)}2^{-n(H(\RV{X} | \RV{U}) - 2\epsilon)}2^{-n(H(\RV{Y} | \RV{W}\RV{U})-2\epsilon)}2^{-n(H(\RV{W}\RV{U})-\epsilon)}\\
      & = 2^{-n(K - 6\epsilon)}
    \end{align} 
    where $K$ is given by
    \begin{align}
      K & = -H(\RV{X}\RV{Y}\RV{W}\RV{U}) +  H(\RV{X} | \RV{U}) + H(\RV{Y} | \RV{W}\RV{U}) + H(\RV{W}\RV{U})\\
      & = \I(\RV{X}; \RV{W} | \RV{U}) + \I(\RV{X}; \RV{Y} | \RV{W}\RV{U})\\
      & \stackrel{(a)}{=} \I(\RV{X}; \RV{Y} \RV{W} | \RV{U})
    \end{align}
    and $(a)$ follows from the chain rule.
  \end{proof}
  \begin{corollary}\label{corollary:lemma:sdd:mixed.strategy:2}
    Note that if the Markov condition $\RV{X}\Markov \RV{U}\Markov \RV{W}$ holds then
    \begin{equation}
      \I\left(\RV{X}; \RV{Y}, \RV{W} | \RV{U}\right) = \I\left(\RV{X}; \RV{Y} | \RV{W}, \RV{U}\right),
    \end{equation}
    as an immediate consequence of the chain rule for mutual information.
  \end{corollary}
  \subsection{Random coding}
    We consider in the following $N+1$ codebooks $C_i, i\in[0; N]$, where in block $b$ codebook $C_{\mathrm{mod}(b, N+1)}$
    is used. The sequences for each codebook are generated as described in the following.
    \begin{itemize}
      \item 
	\textbf{Relay level $\mathbf{N}$} creates the sequences $\RV{v}_N^k\left(q_N^k | q_N^{[1;k-1]}\right), k\in[1;N], q_N^k\in[1; 2^{nR_s^k}],$ which are drawn from
	$\RV{V}_N^k \drawniid p\left(\RV{v}_N^k | \RV{\mathbf{v}}_N^{i\in[1;k-1]}\left(q_N^i | q_N^{[1;i-1]}\right)\right)$ for each
	$q_N^{[1;k-1]}$. Then it creates
	the broadcast messages $\RV{w}_N^1\left(z_N^1 | q_N^{[1;N]}\right)$, $z_N^1\in[1; 2^{n\hat R_N^1}]$, drawn from
	\begin{equation}
	  \RV{W}_N^1 \drawniid p\left(w_N^1 | \RV{\mathbf{v}}_N^{i\in[1;N]}\left(q_N^i | q_N^{[1;i-1]}\right)\right),
	\end{equation}
	where $\RV{\mathbf{v}}_N^{i\in[1;k-1]}\left(q_N^i | q_N^{[1;i-1]}\right)$ abbreviates the set $\left\{\RV{v}_N^i\left(q_N^i | q_N^{[1;i-1]}\right): i\in[1;k-1]\right\}$.
	The sequences are always generated in an ascending order, i.\,e., starting with all $q_N^1\in[1; 2^{nR_s^1}]$ and proceeding
	with increasing $k$.
	Furthermore, the quantization messages $\RV{\hat y}_N^1\left(r_N^1 \right| \left. q_s^{[1;N]}, \left\{q_{[i;N]}^i, z_i^1: i\in[1; N]\right\}\right)$,
	$r_N^1\in\left[1; 2^{n\Delta_N^1}\right]$, are drawn from
	\begin{align}
	  \begin{split}
	    \RV{\hat Y}_N^1\drawniid &
	    p\Bigl(\RV{\hat y}_N^1 | \RV{\mathbf{u}}_s^{m\in[1;N]}\left(q_s^m | q_s^{[1;m-1]}, \left\{q_{[i;N]}^i: i\in\left[1;m\right]\right\}\right),\\
	    & \left\{\RV{\mathbf{v}}_m^{i\in[1;m]}\left(q_m^i | q_m^{[1;i-1]}, q_{[m+1;N]}^{[1;i]}\right), \RV{w}_m^1\left(z_m^1 | q_{[m;N]}^{[1;m]}\right) : m\in[1;N]\right\}\Bigr),
	  \end{split}
	\end{align}
	for each $q_s^{[1;N]}, \left\{q_{[i;N]}^i, z_i^1: i\in[1; N]\right\}$.
	We further use the following \emph{random partitioning}: Each $r_N^1\in[1; 2^{n\Delta_N^1}]$ is randomly assigned to one of the cells
	$Z_N^1(z_N^1)$, $z_N^1\in[1; 2^{n\hat R_N^1}]$, according to a uniform distribution. 

      \item
	\textbf{Relay level $\bm{l}$}, $l\in[1;N]$, creates the supporting messages $\RV{v}_l^k\left(q_l^k | q_l^{[1;k-1]}, q_{[l+1;N]}^{[1;k]}\right)$,
	$k\in[1;l], q_l^k\in\left[1; 2^{nR_s^k}\right]$, drawn from
	\begin{equation}
	  \RV{V}_l^k \drawniid p\left(\RV{v}_l^k | 
	  \RV{\mathbf{v}}_l^{i\in[1;k-1]}\left(q_l^i | q_l^{[1;i-1]}, q_{[l+1;N]}^{[1;i]}\right),
	  \RV{\mathbf{v}}_{i\in[l+1;N]}^k\left(q_{i}^k | q_{i}^{[1;k-1]}, q_{[i+1;N]}^{[1;k]}\right)\right),
	\end{equation}
	for each $q_l^{[1;k-1]}, q_{[l+1;N]}^{[1;k]}$.
	In the next step we create the broadcast sequences $\RV{w}_l^k\left(z_l^k | z_l^{[1;k-1]}, q_{[l;N]}^{[1;l]}\right)$, $k\in[1;M_l]$, $z_l^k\in\left[1;2^{n\hat R_l^k}\right]$,
	which are drawn from
	\begin{equation}
	  \RV{W}_l^k \drawniid p\left(\RV{w}_l^k | \RV{\mathbf{w}}_l^{i\in[1;k-1]}\left(z_l^{i} | z_l^{[1;i-1]}, q_{[l;N]}^{[1;l]}\right)
	  \RV{\mathbf{v}}_l^{i\in[1;l]}\left(q_l^i | q_l^{[1;i-1]}, q_{[l+1;N]}^{[1;i]}\right)\right),
	\end{equation}
	for each $z_l^{[1;k-1]}, q_{[l;N]}^{[1;l]}$.
	We further create the quantizations
	$\RV{\hat y}_l^k\left(r_l^k \right| \left. r_l^{[1;k-1]}, q_s^{[1;l]}, \left\{q_{[i;N]}^i, z_i^1: i\in[1; l]\right\}\right)$,
	$k\in[1;M_l], r_l^k\in\left[1; 2^{n\left(\Delta_l^k-\Delta_l^{k-1}\right)}\right]$, which are drawn from
	\begin{align}
	  \begin{split}
	    \RV{\hat Y}_l^k\drawniid &
	    p\Bigl(\RV{\hat y}_l^k | \RV{\mathbf{u}}_s^{m\in[1;l]}\left(q_s^m | q_s^{[1;m-1]}, \left\{q_{[i;N]}^i: i\in\left[1;m\right]\right\}\right),\\
	    & \RV{\hat{y}}_l^{k-1}\left(r_l^{k-1} | r_l^{[1;k-2]}, q_s^{[1;l]}, \left\{q_{[i;N]}^i, z_i^1: i\in[1; l]\right\}\right),\\
	    & \left\{\RV{\mathbf{v}}_m^{i\in[1;m]}\left(q_m^i | q_m^{[1;i-1]}, q_{[m+1;N]}^{[1;i]}\right), \RV{w}_m^1\left(z_m^1 | q_{[m;N]}^{[1;m]}\right) : m\in[1;l]\right\},\\
	    & \left\{\RV{\mathbf{v}}_m^{i\in[1;l]}\left(q_m^i | q_m^{[1;i-1]}, q_{[m+1;N]}^{[1;i]}\right): m\in[l+1;N]\right\} \Bigr),
	  \end{split}
	\end{align}
	for each $r_l^{[1;k-1]}, q_s^{[1;l]}, \left\{q_{[i;N]}^i, z_i^1: i\in[1; l]\right\}$.
	The joint pdf in (\ref{eq:sdd.mixed:132}) as well as the generation of quantization messages shows
	the Markov chain $\RV{\hat Y}_l^{M_l}\Markov \RV{\hat Y}_l^{M_l-1}\Markov\cdots \RV{\hat Y}_l^1$ which is used in the
	sequel in the context of the successive refinement problem.
	We further use the following \emph{random partitioning}: Each $r_l^i\in[1; 2^{n(\Delta_l^i - \Delta_l^{i-1})}]$ is randomly assigned to one of the cells
	$Z_l^i(z_l^i)$, $z_l^i\in[1; 2^{n\hat R_l^i}]$, according to a uniform distribution. 

      \item \textbf{The source} generates the sequences 
	$u_s^k\left(q_s^k | q_s^{[1;k-1]}, \left\{q_{[i;N]}^i: i\in[1;k]\right\}\right)$ with $q_s^k\in[1; 2^{nR_s^k}]$, $k\in[1;N+1]$, drawn from 
	\begin{equation}
	    \RV{U}_s^k\drawniid p\Bigl(\RV{u}_s^k | \RV{\mathbf{u}}_s^{m\in[1;k-1]}\left(q_s^m | q_s^{[1;m-1]}, \left\{q_{[i;N]}^i: i\in[1;m]\right\}\right),
	    \RV{\mathbf{v}}_{i\in[k;N]}^k\left(q_i^k | q_i^{[1;k-1]}, q_{[i+1;N]}^{[1;k]}\right)\Bigr)
	\end{equation}
	for each $q_s^{[1;k-1]}$ and $\left\{q_{[i;N]}^i: i\in[1;k]\right\}$.
    \end{itemize}
    Note that all codebooks are revealed to all nodes in the network.

  \subsection{Encoding}
    Note that $q_{s,[B-N+1; B]}^{[1;N+1]}=0$, $q_{l, [1;l]}^{[1;l]} = 0$ and $q_{l,[B-N+l+1; B]}^{[1;l]}=0$ which is known to all nodes. 
    Consider now the encoding for block $b$:
    \begin{itemize}
      \item \textbf{At the source node}

	In block $b$ the source transmits the message indices $\left(q_{s,b}^1, \cdots, q_{s,b}^{N+1}\right)$.
	Assume all relays correctly decoded the previous $N$ source transmissions, then 
	the source knows the indices $q_{k\in[1;N],b}^{[1;k]}=q_{s,b-k}^{[1;k]}$
	(and henceforth also the corresponding messages). Therefore, the source transmits the messages
	\begin{equation}
	  \left\{\RV{u}_s^l\left(q_{s,b}^l | q_{s,b}^{[1;l-1]}, \left\{q_{[k;N],b}^k: k\in[1;l]\right\}\right): l\in[1;N]\right\}
	\end{equation}
	from codebook $C_{\mathrm{mod}(b, N+1)}$.

      \item \textbf{At relay level $\bm{l}$}

	Assume the relay successfully decoded $\left\{q_{s,b-l'}^{[1;l]}: l'\in[l;N]\right\}$. 
	In block $b$ level $l$ is supposed to transmit the indices $q_{l,b}^{[1;l]}=q_{s,b-l}^{[1;l]}$.
	It further knows the indices
	transmitted by the subsequent relays $q_{l'\in[l+1;N],b}^{[1;l]}$ for the first $l$ source message levels and can therefore
	support the transmission of these relays.

	Furthermore, let the relay found the quantization indices $\left(r_{l,b-1}^1, \cdots, r_{l,b-1}^{M_l}\right)$ (we define in the
	decoding section how these indices are found). With $r_{l,b-1}^i\in Z_l^i(z_{l,b}^i), i\in[1;M_l]$, the node can create the messages
	$\RV{w}_l^i(z_{l,b}^i | z_{l,b}^{[1;i-1]}, q_{[l;N],b}^{[1;l]}), i\in[1;M_l]$. 
	The relay finally transmits $\RV{v}_l^{k\in[1;l]}\left(q_{l,b}^k | q_{l,b}^{[1;k-1]}, q_{[l+1;N],b}^{[1;k]}\right)$
	and $\RV{w}_l^{k\in[1;M_l]}\left(z_{l,b}^k | z_{l,b}^{[1;k-1]}, q_{[l;N],b}^{[1;l]}\right)$.
    \end{itemize}

  \subsection{Decoding}
    The decoding is described for some arbitrary node $l\in[1;N+1]$, source message level $k$ and block $b$.
    \begin{itemize}
      \item At first we decode the quantizations communicated by
	level $k-1$ (for $k>1$), i.\,e., 
	the broadcast message indices $z_{k-1, b-l+k}^{[1;\inverseOrder_{k-1}(l)]}$ and the corresponding
	quantization indices $r_{k-1, b-l+k-1}^{[1; \inverseOrder_{k-1}(l)]}$. At first we decode the
	broadcast message indices $z_{k-1, b-l+k}^{[1;\inverseOrder_{k-1}(l)]}$. Then, for each
	$j\in[1; \inverseOrder_{k-1}(l)]$, we build the set 
	\begin{equation}
	  \begin{split}
	    \Set{Q}_{l,k-1,b}^{j} = & \biggl\{ \tilde r_{k-1, b-l+k-1}^{j}:\Bigl(\RV{y}_l(b-l+1), \\
	    &   \RV{\hat y}_{k-1}^{j}\left(\tilde r_{k-1, b-l+k-1}^{j} | r_{k-1, b-l+k-1}^{[1;{j}-1]}, q_{s, b-l+1}^{[1;k-1]}, \left\{q_{[i,N], b-l+1}^i, z_i^1: i\in[1;k-1]\right\}\right),\\
	    &   \RV{\hat y}_{k-1}^{{j}-1}\left(r_{k-1, b-l+k-1}^{{j}-1} | r_{k-1, b-l+k-1}^{[1;{j}-2]}, q_{s, b-l+1}^{[1;k-1]}, \left\{q_{[i,N], b-l+1}^i, z_{i, b-l+1}^1: i\in[1;k-1]\right\}\right),\\
	    &   \RV{\mathbf{\hat y}}_{i\in[1;k-2]}^{\inverseOrder_i(l)}\left(r_{i, b-l+i}^{\inverseOrder_i(l)} | r_{i, b-l+i}^{[1;\inverseOrder_i(l)-1]}, q_{s, b-l+1}^{[1;i]}, \left\{q_{[m,N], b-l+1}^m, z_{m, b-l+1}^1: m\in[1;i]\right\}\right),\\
	    &   \RV{\mathbf{u}}_s^{[1;k-1]}, \left\{\RV{\mathbf{v}}_i^{m\in[1;i]}, \RV{\mathbf{w}}_i^{m\in[1;\inverseOrder_i(l)]}: i\in[1;l]\right\},
		\RV{\mathbf{v}}_{i\in[l+1;N]}^{[1;l]}
	      \Bigr)\in\StronglyTypicalSet
	      \biggr\},
	  \end{split}\label{eq:appendix:proof.full-duplex-mixed:250}
	\end{equation}
	where we dropped the block index $b-l+1$ if it is clear from the context. The set $\Set{Q}_{l,k-1,b}^{j}$ holds
	all possible quantization indices $\tilde r_{k-1, b-l+k-1}^j$ and we search for such an unique index
	which maps to $Z_{k-1}^j(z_{k-1, b-l+k}^j)$ as follows
	\begin{equation}
	  \exists \tilde r_{k-1,b-l+k-1}^{j}: \tilde r_{k-1,b-l+k-1}^{j} = \Set{Q}_{l,k-1,b}^{j}\cap \Set{Z}_{k-1}^{j}(z_{k-1,b-l+k}^{j}).\label{eq:appendix:proof.full-duplex-mixed:251}
	\end{equation}
	\begin{remark}
	  We must note that when decoding source message level $k$ we can only decode the indices of nodes $[k;l]$. Furthermore, we
	  decode in this step only the quantizations of node level $k-1$ which exploits the quantization information
	  acquired in the previous $k-2$ decoding steps (in the first step we do not decode any quantization information).
	\end{remark}

      \item Using these quantization messages we proceed with the decoding of the source message $q_{s,b-l+1}^k$. 
	Consider at first the following sets:
	\begin{align}
	  \begin{split}
	    \Set{T}_{l,b,j}^k = & \biggl\{
	    \tilde q_{j,b-l+1+j}^k: \biggl(\RV{v}_j^k\left(\tilde q_{j,b-l+1+j}^k | q_{j,b-l+1+j}^{[1;k-1]}, q_{[j+1;N], b-l+1+j}^{[1;k]}\right),
	    \RV{y}_l(b-l+1+j)\\
	    & \RV{v}_j^{[1;k-1]}, \left\{\RV{v}_i^{[1;i]}, \RV{w}_i^{[1;\inverseOrder_i(l)]}: i\in[j+1;l]\right\}, \RV{v}_{[l;N]}^{[1;l]}
	    \biggr)\in\StronglyTypicalSet\biggr\} 
	  \end{split}\label{eq:sdd.mixed:100}\\
	  \begin{split}
	    \Set{T}_{l,b,0}^k = & \biggl\{
	    \tilde q_{s,b-l+1}^k: \biggl(
	    \RV{u}_s^k\left(\tilde q_{s,b-l+1}^k | q_{s,b-l+1}^{[1;k-1]}, \left\{q_{[i;N],b-l+1}^i: i\in[1;k]\right\}\right),
	    \RV{y}_l(b-l+1)\\
	    & \RV{\hat y}_{l'\in[1;k-1]}^{\inverseOrder_{l'}(l)}\left(r_{l', b-l+l'}^{\inverseOrder_{l'}(l)} | 
	    r_{l', b-l+l'}^{[1; \inverseOrder_{l'}(l)-1]}, q_{s,b-l+1}^{[1;l']}, 
	    \left\{ q_{[j;N], b-l+1}^j, z_{j,b-l+1}^1: j\in[1;l'] \right\}\right) \\
	    & \RV{u}_s^{[1;k-1]}, \left\{\RV{v}_{[i;N]}^i, \RV{w}_i^{[1; \inverseOrder_i(l)]}: i\in[1;l]\right\}\biggr)\in\StronglyTypicalSet
	    \biggr\} 
	  \end{split}\label{eq:sdd.mixed:101}
	\end{align}
	for all $j\in[k;l-1]$. Note that we dropped in (\ref{eq:sdd.mixed:100}) and (\ref{eq:sdd.mixed:101}) the block index $b-l+1+j$
	where it is uniquely defined by the context. We finally decode $q_{s,b-l+1}^k$ iff
	\begin{equation}
	  \exists \tilde q_{s,b-l+1}^k: \tilde q_{s,b-l+1}^k = \Set{T}_{l,b,0}^k \cap \bigcap\limits_{j=k}^{l-1} \Set{T}_{l,b,j}^k.\label{eq:sdd.mixed:102}
	\end{equation}

      \item
	Finally, if $l<N+1$, the relay needs to quantize its channel output in block $b-l+1$ after decoding all source message indices
	$q_{s,b-l+1}^{[1;l]}$. This is done for a suitable quantization index $\tilde r_{l,b}^j$, $j\in[1;M_l]$, such that
	\begin{equation}
	  \begin{split}
	    \exists \tilde r_{l, b}^j: & \biggl(
	      \RV{\hat y}_l^j\left(\tilde r_{l, b}^j | r_{l, b}^{[1;j-1]}, q_{s,b-l+1}^{[1;l]}, \left\{s_{[i;N], b-l+1}^i, z_{i,b-l+1}^1: i\in[1;l]\right\}\right),\\
	      & \RV{\hat y}_l^{j-1}\left(r_{l, b}^{j-1} | r_{l, b}^{[1;j-2]}, q_{s,b-l+1}^{[1;l]}, \left\{s_{[i;N], b-l+1}^i, z_{i,b-l+1}^1: i\in[1;l]\right\}\right),
	      \RV{y}_l(b-l+1),\\
	      & \RV{\mathbf{u}}_s^{j\in[1;l]}, \left\{\RV{\mathbf{v}}_i^{j\in[1;i]}, \RV{\mathbf{w}}_i^{j\in[1;\inverseOrder_i(l)]}: i\in[1;l]\right\}, \RV{\mathbf{v}}_{i\in[l;N]}^{[1;l]}
	    \biggr)\in\StronglyTypicalSet,
	  \end{split}\label{eq:sdd.mixed:103}
	\end{equation}
	where we again dropped the block index $b-l+1$ if it is uniquely defined by the context.

    \end{itemize}

  \subsection{Definition of error events}
    Consider the following error events which might occur using the previous decoding rule:
    \begin{itemize}
      \item $E_{0,b}$: the error that all previously described r.v.s are not $\epsilon$-jointly typical in block $b$,
      \item $E_{l,b}^k$: the error that level $l$ does not correctly decode in block $b$
	source message level $k\in[1;l]$,
      \item $E_{l,l',b}^{k'}$: the error that level $l$ does not correctly decode in block $b$
	the broadcast message $k'\in[1;\inverseOrder_{l'}(l)]$ sent by level $l'$,
      \item $\hat E_{l,l',b}^{k'}$: the error that level $l$ does not correctly decode in block $b$
	the quantization message $k'\in[1;\inverseOrder_{l'}(l)]$ sent by level $l'$, and
      \item $\hat E_{l,b}^{k'}$: the error that level $l$ does not find a quantization index $k'$ in block $b$.
    \end{itemize}
    Now let
    \begin{equation}
      F_b=E_{0,b}\cup\bigcup\limits_{l=1}^{N+1}
	\left(
	  \bigcup\limits_{k=1}^{l-1} E_{l,b}^k\cup
	  \bigcup\limits_{l'=1}^{l-1}\bigcup\limits_{k'=1}^{\inverseOrder_{l'}(l)}\left(E_{l,l',b}^{k'}\cup \hat E_{l,l',b}^{k'}\right)\cup
	  \bigcup\limits_{k'=1}^{M_l}\hat E_{l,b}^{k'}
	\right)\label{eq:appendix:proof.full-duplex-mixed:10}
    \end{equation}
    where $\Prob\left\{E_{N+1, b}^{k'}\right\}=0$. 
    We can use $\Prob\left\{F_b | F_{b-1}^c \cup F_{b-2}^c\cup \dots F_{b-N}^c\right\}$ to upper bound the probability of error
    in block $b$. Besides, note that 
    $\Prob\left\{F_b | F_{b-1}^c \cup F_{b-2}^c\cup\dots F_{b-N}^c\right\}=\Prob\left\{F_b | F_{b-1}^c \cup F_{b-1}^c\cup \dots F_{1}^c\right\}$
    as the error in block $b$ only depends on the previous $N$ blocks.

  \subsection{Individual error events probabilities}
    Before proving that the overall probability of error is arbitrarily small, we derive now the individual probabilities
    of the previously listed error events.

    \subsubsection{Event $E_{0,b}$}
      This event is defined by
      \begin{equation}
	\Prob\left\{E_{0,b} | F_{[1;b-1]}^c\right\} = \Prob\left\{\left(\RV{u}_s^{[1;N+1]}, \RV{v}_{i\in[1;N]}^{[1;i]},
	  \RV{w}_{i\in[1;N]}^{[1;M_i]}, \RV{\hat y}_{i\in[1;N]}^{[1;M_i]}, \RV{y}_{[1;N+1]}\right)\notin\StronglyTypicalSet\right\},
      \end{equation}
      where we dropped the block index $b$ to improve the readability.
      From Wyner-Ziv coding \cite{Wyner.Ziv.1976} as well as compress-and-forward \cite[Theorem 6]{Cover.Gamal.TransIT.1979} 
      we know that the proof for this events requires the application
      of the Markov lemma \cite[Lemma 14.8.1]{Cover.Thomas.1991} to ensure joint typicality. 
      From the joint pdf in (\ref{eq:sdd.mixed:132}) we can define the following Markov chain
      \begin{equation}
	\begin{split}
	  & \RV{\hat Y}_l^k\Markov\left(\RV{\hat Y}_l^{k+1}, \RV{U}_s^{[1;l]}, \RV{V}_{m\in[1;l]}^{[1;m]}, \RV{V}_{[l+1;N]}^{[1;l]}, \RV{W}_{[1;l]}^1\right)\Markov \\
	  & \left(\RV{\hat Y}_l^{[k+2;M_l]}, \RV{Y}_{[1;N+1]}, \RV{U}_s^{[l+1;N+1]}, \RV{V}_{l'\in[l+1;N]}^{[l+1;l']}, \RV{W}_{[1;l]}^{[2;M_l]},
	  \RV{W}_{[l+1;N]}^{[1;M_l]}\right)
	\end{split}
      \end{equation}
      where $\RV{\hat Y}_l^{k+1}$ is substituted by $\RV{Y}_l$ if $k=M_l$. After recursive application of this
      Markov chain starting with $k=M_l$ we know using the Markov lemma that 
      \begin{equation}
	\Prob\left\{E_{0,b} | F_{[1;b-1]}^c\right\}\leq\epsilon,\label{eq:appendix:proof.full-duplex-mixed:50}
      \end{equation}
      where $\epsilon\rightarrow 0$ as $n\rightarrow\infty$.

    \subsubsection{Event $E_{l,l',b}^{k'}$}
      When decoding in block $b$ the broadcast message index $z_{l', b-l+l'+1}^{k'}$ we assume that 
      \begin{equation}
	\left(E_{0,b}\cup F_{[b-N;b-1]}\cup E_{l,b}^{[1;l']}\cup E_{l,l',b}^{[1;k'-1]}\right)^c
      \end{equation}
      holds and we therefore know the message indices $q_{[l;N], b-l+l'+1}^{[1;l]}$, $q_{m\in[l'+1;l], b-l+l'+1}^{[1;m]}$,
      $z_{m\in[l'+1;l], b-l+l'+1}^{[1;\inverseOrder_m(l)]}$ and $z_{l', b-l+l'+1}^{[1;k'-1]}$. Using these indices we search for
      \begin{multline}
	\exists \tilde z_{l', b-l+l'+1}^{k'}: \biggl(
	\RV{w}_{l'}^{k'}\left(\tilde z_{l', b-l+l'+1}^{k'} | z_{l', b-l+l'+1}^{[1;k'-1]}, q_{[l';N], b-l+l'+1}^{[1;l']}\right),
	  \RV{y}_l(b-l+l'+1),\\
	  \RV{\mathbf{w}}_{l'}^{i\in[1;k'-1]}, \RV{\mathbf{v}}_{l'}^{i\in[1;l']}, \left\{\RV{\mathbf{v}}_i^{[1;i]}, 
	  \RV{\mathbf{w}}_i^{[1;\inverseOrder_i(l)]}: i\in[l'+1; l]\right\}, \RV{\mathbf{v}}_{i\in[l;N]}^{[1;l]}
	\biggr)\in\StronglyTypicalSet
	\label{eq:appendix:proof.full-duplex-mixed:180}	
      \end{multline}
      where we again dropped the block index $b-l+l'+1$ in the lower part for the sake of readability. 
      If and only if $n$ is sufficient large and 
      \begin{equation}
	\hat R_{l'}^{k'} < \I\left(\RV{W}_{l'}^{k'}; \RV{Y}_l | \RV{V}_{l'}^{[1;l']}, \RV{W}_{l'}^{[1;k'-1]}, 
	  \left\{\RV{V}_i^{[1;i]}, \RV{W}_i^{[1;\inverseOrder_i(l)]}: i\in[l'+1;l]\right\},
	  \RV{V}_{[l+1;N]}^{[1;l]}\right),\label{eq:appendix:proof.full-duplex-mixed:200}
      \end{equation}
      we can state that
      \begin{multline}
	\Prob\left\{E_{l,l',b}^{k'} | \left(E_{0,b}\cup F_{[1;b-1]}\cup E_{l,b}^{[1;l']}\cup E_{l,l',b}^{[1;k'-1]}\right)^c\right\}=\\
	\Prob\left\{\tilde z_{l', b-l+l'+1}^{k'}\neq z_{l', b-l+l'+1}^{k'}\vee\text{ (\ref{eq:appendix:proof.full-duplex-mixed:180}) is not satisfied}\right\}
	\leq\epsilon,\label{eq:appendix:proof.full-duplex-mixed:100}
      \end{multline}
      and $\epsilon\rightarrow0$. Eq. (\ref{eq:appendix:proof.full-duplex-mixed:100}) follows from
      Lemma \ref{lemma:sdd:mixed.strategy:2} and Corollary \ref{corollary:lemma:sdd:mixed.strategy:2} with substitutions
      \begin{align*}
	\RV{X} & \mapsto \RV{W}_{l'}^{k'} & \RV{U} & \mapsto \left(\RV{W}_{l'}^{[1;k'-1]}, \RV{V}_{[l'; N]}^{[1;l']}\right) \\
	\RV{Y} & \mapsto \RV{Y}_l & \RV{W} & \mapsto \left(\RV{\mathbf{V}}_{i\in[l'+1;l-1]}^{[l'+1;i]}, \RV{V}_{[l;N]}^{[l'+1;l]}, \RV{\mathbf{W}}_{i\in[l'+1;l-1]}^{[1;\inverseOrder_i(l)]}, \RV{W}_l^{[1;M_l]}\right),
      \end{align*}
      as well as the fact that $F_{b-l+l'+1}^c$ implies $\tilde z_{l', b-l+l'+1}^{k'}=z_{l', b-l+l'+1}^{k'}$ if there is such a unique $\tilde z_{l', b-l+l'+1}^{k'}$.

    \subsubsection{Event $\hat E_{l,l',b}^{k'}$}
      When decoding $r_{l', b-l+l'}^{k'}$ we assume that
      \begin{equation}
	\left(E_{0,b}\cup F_{[b-N; b-1]} \cup E_{l,l',b}^{[1; k']}\cup \hat E_{l,l',b}^{[1;k'-1]}\cup E_{l,b}^{[1;l']}\cup 
	  \hat E_{l,j'\in[1;l'-1],b}^{[1;\inverseOrder_{j'}(l)]}\cup E_{l,j'\in[1;l'-1],b}^{[1;\inverseOrder_{j'}(l)]}\right)^c
      \end{equation}
      holds, which implies that we know $z_{l', b-l+l'+1}^{[1; k']}$, $r_{l', b-l+l'}^{[1;k'-1]}$, $r_{j'\in[1;l'-1], b-l+j'}^{[1; \inverseOrder_{j'}(l)]}$, $q_{s, b-l+1}^{[1;l']}$, 
      $z_{j'\in[1; l], b-l+1}^{[1;\inverseOrder_{j'}(l)]}$, $q_{j'\in[1;l], b-l+1}^{[1;j']}$ and $q_{[l+1;N], b-l+1}^{[1;l]}$.
      Using these message indices we build the intersection of $\Set{Q}_{l,l',b}^{k'}$ and $\Set{Z}_{l'}^{k'}(z_{l',b-l+l'+1}^{k'})$
      as defined in (\ref{eq:appendix:proof.full-duplex-mixed:250}) and (\ref{eq:appendix:proof.full-duplex-mixed:251}). 
      From $F_{[b-N; b-1]}^c$ we know that the correct quantization index is element of this set. 
      The probability that (\ref{eq:appendix:proof.full-duplex-mixed:251}) is satisfied can be given by
      \begin{equation}
	\Prob\left\{\hat E_{l,l',b}^{k'} | \left(E_{0,b}\cup F_{[b-N; b-1]} \cup E_{l,l',b}^{[1; k']}\cup \hat E_{l,l',b}^{[1;k'-1]}\cup E_{l,b}^{[1;l']}\cup 
	  \hat E_{l,j'\in[1;l'-1],b}^{[1;\inverseOrder_{j'}(l)]}\cup E_{l,j'\in[1;l'-1],b}^{[1;\inverseOrder_{j'}(l)]}\right)^c\right\}\leq\epsilon,\label{eq:appendix:proof.full-duplex-mixed:150}
      \end{equation}
      where $\epsilon\rightarrow0$ iff $n$ is sufficiently large and
      \begin{align}
	\begin{split}
	  \Delta_{l'}^{k'} - \Delta_{l'}^{{k'}-1} & < \hat R_{l'}^{k'} + \I\Bigl(\RV{\hat Y}_{l'}^{k'}; \RV{Y}_l, \RV{\mathbf{\hat Y}}_{i\in[1;k-2]}^{\inverseOrder_i(l)},
	    \left\{\RV{V}_{[i;N]}^i, \RV{W}_i^{[1;\inverseOrder_i(l)]}: i\in[k;l]\right\},\\
	    & \quad\quad\RV{\mathbf{W}}_{i\in[1;l']}^{[2;\inverseOrder_i(l)]} \Bigl\lvert \RV{\hat Y}_{l'}^{{k'}-1}, \RV{U}_s^{[1;l']}, \left\{\RV{V}_{[i;N]}^i, \RV{W}_i^1: i\in[1;l']\right\}\Bigr)
	\end{split}\label{eq:appendix:proof.full-duplex-mixed:252}
      \end{align}
      with $\Delta_{l'}^0 = 0$. We applied in (\ref{eq:appendix:proof.full-duplex-mixed:252}) Lemma \ref{lemma:sdd:mixed.strategy:2} with substitutions
      \begin{align*}
	\RV{X} & \mapsto \RV{\hat Y}_{l'}^{k'} & 
	\RV{U} & \mapsto \left(\RV{\hat Y}_{l'}^{[1;{k'}-1]}, \RV{U}_s^{[1;l']}, \left\{\RV{V}_{[i;N]}^i, \RV{W}_i^1: i\in[1;l']\right\}\right)\\
	\RV{Y} & \mapsto \left(\RV{Y}_l, \RV{\mathbf{\hat Y}}_{i\in[1;k-2]}^{\inverseOrder_i(l)}\right) &
	\RV{W} & \mapsto \left\{\left\{\RV{V}_i^{[1;i]}, \RV{W}_i^{[1;\inverseOrder_i(l)]}: i\in[1;l]\right\}, \RV{V}_{[l+1;N]}^{[1;l]}\right\}\setminus\RV{U},
      \end{align*}
      Using the Markov structure of the quantization messages and the chain rule of mutual
      information we can build the sum over the quantization rates to get
      \begin{align}
	\begin{split}
	  \Delta_{l'}^{k'} & < \I\Bigl(\RV{\hat Y}_{l'}^{k'}; \RV{Y}_l, \RV{\mathbf{\hat Y}}_{i\in[1;k-2]}^{\inverseOrder_i(l)},
	    \left\{\RV{V}_{[i;N]}^i, \RV{W}_i^{[1;\inverseOrder_i(l)]}: i\in[k;l]\right\},\\
	    & \quad\quad\RV{\mathbf{W}}_{i\in[1;l']}^{[2;\inverseOrder_i(l)]} | \RV{U}_s^{[1;l']}, \left\{\RV{V}_{[i;N]}^i, \RV{W}_i^1: i\in[1;l']\right\}\Bigr)
	    + \sum\limits_{i=1}^{k'} \hat R_{l'}^i
	\end{split}.\label{eq:appendix:proof.full-duplex-mixed:260}
      \end{align}

    \subsubsection{Event $E_{l,b}^k$}
      When decoding the source message index $q_{s,b-l+1}^k$, we assume that
      \begin{equation}
	\left(E_{0,b}\cup F_{[b-N;b-1]}\cup E_{l,b}^{[1;k-1]}\cup \hat E_{l,l'\in[1;k-1],b}^{[1;\inverseOrder_{l'}(l)]}\cup E_{l,l'\in[1;k-1],b}^{[1;\inverseOrder_{l'}(l)]}\right)^c
      \end{equation}
      holds, which implies that the indices $q_{j\in\left\{0, [k;l-1]\right\}, b-l+1+j}^{[1;k-1]}$, $z_{j'\in[j+1; l], b-l+1+j}^{[1;\inverseOrder_{j'}(l)]}$,
      $q_{j'\in[j+1;l], b-l+1+j}^{[1;j']}$, $q_{[l+1;N], b-l+1+j}^{[1;l]}$ and $r_{l'\in[1;k-1], b-l+l'}^{[1; \inverseOrder_{l'}(l)]}$ are known.
      According to (\ref{eq:sdd.mixed:102}) we must prove that the correct index $q_{s,b-l+1}^k$ is the only element of the intersection
      \begin{equation}
	\Set{T}_{l,b}^k=\Set{T}_{l,b,0}^k\cap\bigcap\limits_{j=k}^{l-1}\Set{T}_{l,b,j}^k,
      \end{equation}
      where $\Set{T}_{l,b,0}^k$ and $\Set{T}_{l,b,[k;l-1]}^k$ are defined in (\ref{eq:sdd.mixed:100}) and (\ref{eq:sdd.mixed:101}). 
      From $F_{[b-N;b-1]}^c$ we know that the correct index will be element of each these sets. 
      Using Lemma \ref{lemma:sdd:mixed.strategy:2} the probability that a wrong index is element of one these sets is
      \begin{multline}
	\Prob\left\{\tilde q_{j,b-l+1+j}^k\neq q_{j,b-l+1+j}^k \wedge \tilde q_{j,b-l+1+j}^k\in\Set{T}_{l,b,j}^k\right\} \doteq\\
	2^{-n\left(\I\left(\RV{V}_j^k; \RV{Y}_l | \RV{V}_j^{[1;k-1]},
	\left\{\RV{V}_i^{[1;i]}, \RV{W}_i^{[1;\inverseOrder_i(l)]} : i\in[j+1;l]\right\}, \RV{\mathbf{V}}_{[l;N]}^{[1;l]}\right)\right)}
      \end{multline}
      and
      \begin{multline}
	\Prob\left\{\tilde q_{s,b-l+1}^k\neq q_{s,b-l+1}^k \wedge \tilde q_{s,b-l+1}^k\in\Set{T}_{l,b,0}^k\right\} \doteq
	2^{-n\left(\I\left(\RV{U}_s^k; \RV{Y}_l, \RV{\hat Y}_{l'\in[1;k-1]}^{\inverseOrder_{l'}(l)} | \RV{U}_s^{[1;k-1]}, 
	\left\{\RV{V}_{[i;N]}^i, \RV{W}_i^{[1;\inverseOrder_i(l)]}: i\in[1;l]\right\}\right)\right)}
      \end{multline}
      where $q_{s,b-l+1}^k=q_{j,b-l+1+j}^k$ as $F_{[b-N; b-1]}^c$ holds.
      From the independence of the used codebooks we know that
      \begin{equation*}
	\begin{split}
	  \Prob\left\{\tilde q_{s,b-l+1}^k\neq q_{s,b-l+1}^k \wedge \tilde q_{s,b-l+1}^k\in\Set{T}_{l,b}^k\right\} = 
	  \Prob\left\{\tilde q_{s,b-l+1}^k\neq q_{s,b-l+1}^k \wedge \tilde q_{s,b-l+1}^k\in\Set{T}_{l,b,0}^k\right\}\\
	  \cdot\prod\limits_{j=k}^{l-1} \Prob\left\{\tilde q_{s,b-l+1}^k\neq q_{s,b-l+1}^k \wedge \tilde q_{s,b-l+1}^k\in\Set{T}_{l,b,j}^k\right\}.
	\end{split}
      \end{equation*}
      As there are $2^{nR_s^k} - 1$ possible wrong indices it follows that
      \begin{equation}
	\begin{split}
	  R_s^k < & \I\left(\RV{U}_s^k; \RV{Y}_l, \RV{\hat Y}_{l'\in[1;k-1]}^{\inverseOrder_{l'}(l)} | \RV{U}_s^{[1;k-1]}, 
	  \left\{\RV{V}_{[i;N]}^i, \RV{W}_i^{[1;\inverseOrder_i(l)]}: i\in[1;l]\right\}\right) + \\
	  & \,\sum\limits_{j=k}^{l-1} \I\left(\RV{V}_j^k; \RV{Y}_l | \RV{V}_j^{[1;k-1]},
	  \left\{\RV{V}_i^{[1;i]}, \RV{W}_i^{[1;\inverseOrder_i(l)]} : i\in[j+1;l]\right\}, \RV{\mathbf{V}}_{[l;N]}^{[1;l]}\right)
	\end{split}\label{eq:appendix:proof.full-duplex-mixed:300}
      \end{equation}
      must hold and $n\rightarrow\infty$ such that
      \begin{equation}
	\Prob\left\{E_{l,b}^k | \left(E_{0,b}\cup F_{[1;b-1]}\cup E_{l,b}^{[1;k-1]}\cup \hat E_{l,l'\in[1;k-1],b}^{[1;\inverseOrder_{l'}(l)]}\cup E_{l,l'\in[1;k-1],b}^{[1;\inverseOrder_{l'}(l)]}\right)^c\right\}
	\leq\epsilon,\label{eq:appendix:proof.full-duplex-mixed:350}
      \end{equation}
      with $\epsilon\rightarrow0$.
      
    \subsubsection{Event $\hat E_{l,b}^k$}
      Before we build the quantization index $r_{l,b}^k$ we assume
      \begin{equation}
	\left(E_{0,b}\cup F_{[b-N; b-1]}\cup E_{l,b}^{[1;l]}\cup E_{l,b}^{[1;k-1]}\right)^c
      \end{equation}
      which implies that $q_{s,b-l+1}^{[1;l]}$, $z_{[1;l], b-l+1}^1$, $q_{l'\in[1;l],b-l+1}^{[1;l']}$, $q_{l'\in[l+1;N], b-l+1}^{[1;l]}$ and $r_{l,b}^{[1;k-1]}$.
      are known.  An error during the quantization occurs if (\ref{eq:sdd.mixed:103}) cannot be satisfied, hence we can state that
      \begin{equation}
	\Prob\left\{\hat E_{l,b}^k | F_{[1;b-1]}^c\cap \hat E_{l,b}^{[1;k-1],c}\cap E_{l,b}^{[1;l],c}\cap 
	\hat E_{l,j'\in[1;l-1],b}^{[1;\inverseOrder_{j'}(l)],c}\cap E_{l,j'\in[1;l-1],b}^{[1;\inverseOrder_{j'}(l)],c}\right\}
	\leq\epsilon.\label{eq:appendix:proof.full-duplex-mixed:405}
      \end{equation}
      with $\epsilon\rightarrow0$ iff $n\rightarrow\infty$ and
      \begin{eqnarray}
	\Delta_l^j - \Delta_l^{j-1} & > & \I\left(\RV{\hat Y}_l^{j}; \RV{Y}_l | \RV{\hat Y}_l^{j-1}, \RV{U}_s^{[1;l]}, 
	  \left\{\RV{V}_{[i;N]}^i, \RV{W}_i^1: i\in[1;l]\right\}\right)\\
	\Delta_l^j & > & \I\left(\RV{\hat Y}_l^{j}; \RV{Y}_l | \RV{U}_s^{[1;l]}, 
	  \left\{\RV{V}_{[i;N]}^i, \RV{W}_i^1: i\in[1;l]\right\}\right)\label{eq:appendix:proof.full-duplex-mixed:410}
      \end{eqnarray}
      which results from the rate distortion theorem \cite[Theorem 13.2.1]{Cover.Thomas.1991}, the chain rule for mutual information as well as the
      previously discussed Markov structure within the quantizations.
      Using (\ref{eq:appendix:proof.full-duplex-mixed:260}) we can state for the quantization source coding rates
      \begin{multline}
	\sum\limits_{i=1}^m \hat R_l^i = R_{\RV{Y}_l | \cdot}^{\text{WZ}}(D_l^m) >
	\max\limits_{\substack{j\in[m;M_l]\\ j'=o_l(j)}} \biggl[\Delta_l^m -
	\I\Bigl(\RV{\hat Y}_l^m; \RV{Y}_{j'}, \RV{\mathbf{\hat Y}}_{i\in[1;l-1]}^{\inverseOrder_i(j')},
	\left\{\RV{V}_{[i;N]}^i, \RV{W}_i^{[1;\inverseOrder_i(l)]}: i\in[l+1;j']\right\},\\
	\RV{\mathbf{W}}_{i\in[1;l]}^{[2;\inverseOrder_i(j')]} | \RV{U}_s^{[1;l]}, \left\{\RV{V}_{[i;N]}^i, \RV{W}_i^1: i\in[1;l]\right\}\biggr], m\in[1; M_l].\label{eq:appendix:proof.full-duplex-mixed:411}
      \end{multline}
      If we use (\ref{eq:appendix:proof.full-duplex-mixed:410}) in (\ref{eq:appendix:proof.full-duplex-mixed:411}) it follows that
      \begin{multline}
	\sum\limits_{i=1}^m \hat R_l^i = \max\limits_{\substack{j\in[m;M_l]\\ j'=o_l(j)}} \biggl[ \I\left(\RV{\hat Y}_l^m; \RV{Y}_l | \RV{U}_s^{[1;l]}, 
	\left\{\RV{V}_{[i;N]}^i, \RV{W}_i^1: i\in[1;l]\right\}\right) -
	\I\Bigl(\RV{\hat Y}_l^m; \RV{Y}_{j'}, \RV{\mathbf{\hat Y}}_{i\in[1;l-1]}^{\inverseOrder_i(j')},\\
	\left\{\RV{V}_{[i;N]}^i, \RV{W}_i^{[1;\inverseOrder_i(l)]}: i\in[l+1;j']\right\}, 
	\RV{\mathbf{W}}_{i\in[1;l]}^{[2;\inverseOrder_i(j')]} | \RV{U}_s^{[1;l]}, \left\{\RV{V}_{[i;N]}^i, \RV{W}_i^1: i\in[1;l]\right\}\biggr]\label{eq:sdd.mixed:160}
      \end{multline}
      Consider now the following mapping
      \begin{align*}
	\RV{W} & \mapsto \left\{\RV{V}_{[i;N]}^i, \RV{W}_i^{[1;\inverseOrder_i(j')]}: i\in[l+1;j']\right\}, 
	\RV{\mathbf{W}}_{i\in[1;l]}^{[2;\inverseOrder_i(j')]}, & \RV{Y} & \mapsto \left(\RV{Y}_{j'}, \RV{\mathbf{\hat Y}}_{i\in[1;l-1]}^{\inverseOrder_i(j')}\right)\\
	\RV{U} & \mapsto \left(\RV{U}_s^{[1;l]}, \left\{\RV{V}_{[i;N]}^i, \RV{W}_i^1: i\in[1;l]\right\}\right), & 
	\RV{X} & \mapsto \RV{\hat Y}_{l}^m.
      \end{align*}
      Due to the Markov chain $\RV{\hat X}\Markov \left(\RV{Y}_l, \RV{U}\right)\Markov \left(\RV{Y}, \RV{W}\right)$ it follows that
      $\mathrm{H}\left(\RV{\hat X} | \RV{Y}_l, \RV{U}\right) = \mathrm{H}\left(\RV{\hat X} | \RV{Y}_l, \RV{Y}, \RV{W}, \RV{U}\right)$ which gives
      \begin{align*}
	\I\left(\RV{\hat X}; \RV{Y}_l | \RV{U}\right) - \I\left(\RV{\hat X}; \RV{Y}, \RV{W} | \RV{U}\right) & = 
	\mathrm{H}\left(\RV{\hat X} | \RV{U}\right) - \mathrm{H}\left(\RV{\hat X} | \RV{U}\right) - \mathrm{H}\left(\RV{\hat X} | \RV{Y}_l, \RV{U}\right) + \mathrm{H}\left(\RV{\hat X} | \RV{Y}, \RV{W}, \RV{U}\right)\\
	& = \I\left(\RV{\hat X}; \RV{Y}_l | \RV{Y}, \RV{W}, \RV{U}\right).
      \end{align*}
      Applying this result to (\ref{eq:sdd.mixed:160}) we have
      \begin{equation}
	\sum\limits_{i=1}^m \hat R_l^i = \max\limits_{\substack{j\in[m;M_l]\\ j'=o_l(j)}}
	  \I\Bigl(\RV{\hat Y}_l^m; \RV{Y}_l | \RV{Y}_{j'}, \RV{\mathbf{\hat Y}}_{i\in[1;l-1]}^{\inverseOrder_i(j')},
	  \RV{U}_s^{[1;l]}, \left\{\RV{V}_{[i;N]}^i, \RV{W}_i^{[1;\inverseOrder_i(j')]}: i\in[1;j']\right\}\Bigr).
      \end{equation}

  \subsection{Bringing all together}
    The previous events were necessary to prove the error in block $b$. As we are interested in the overall error probability after $B$ blocks,
    we need to consider the overall error probability $P_e$ using standard techniques:
    \begin{equation}
      P_e \leq \Prob\left\{\bigcup\limits_{b=1}^B F_b\right\}
      \leq \sum\limits_{b=1}^B \Prob\left\{F_b \cap F_{[1; b-1]}^c\right\}.\label{eq:appendix:proof.full-duplex-mixed:505}
    \end{equation}
    Using the definition of $F_b$ in (\ref{eq:appendix:proof.full-duplex-mixed:10}) we get
    \begin{align}
      \Prob\left\{F_b\cap F_{[1;b-1]}^c\right\} = & \Prob\left\{E_{0,b}\cup\bigcup\limits_{l=1}^{N+1}
	\left(
	  \bigcup\limits_{k=1}^{l-1} E_{l,b}^k\cup
	  \bigcup\limits_{l'=1}^{l-1}\bigcup\limits_{k'=1}^{\inverseOrder_{l'}(l)}\left(E_{l,l',b}^{k'}\cup \hat E_{l,l',b}^{k'}\right)\cup
	  \bigcup\limits_{k'=1}^{M_l}\hat E_{l,b}^{k'}
	  \right) \cap F_{[1;b-1]}^c\right\}\\
      \begin{split}	  
	\leq & (\ref{eq:appendix:proof.full-duplex-mixed:50}) +
	\sum\limits_{l=1}^N\left( \sum\limits_{k=1}^{l-1} (\ref{eq:appendix:proof.full-duplex-mixed:350}) + 
	   \sum\limits_{l'=1}^{l-1}\sum\limits_{k'=1}^{\inverseOrder_{l'}(l)}\left[ (\ref{eq:appendix:proof.full-duplex-mixed:100}) + (\ref{eq:appendix:proof.full-duplex-mixed:150}) \right]
	   + \sum\limits_{k'=1}^{M_l} (\ref{eq:appendix:proof.full-duplex-mixed:405})
	\right).
      \end{split}\label{eq:appendix:proof.full-duplex-mixed:510}
    \end{align}
    Using this result in (\ref{eq:appendix:proof.full-duplex-mixed:505}) we have
    \begin{equation}
      P_e \leq B\left(1 + N\left(2N + N^2\right)\right)\epsilon,
    \end{equation}
    which can be made arbitrarily small as $n$ grows to infinity. Now lets apply the standard random coding argument
    by throwing away the worse half of codewords. It follows for the overall rate
    \begin{equation}
      \tilde R = \frac{nR\left(B-N\right)-1}{nB},
    \end{equation}
    which approaches $R$ as $n\rightarrow\infty$ and then $B\rightarrow\infty$. This and the previously given bounds 
    prove that $R$ in Theorem \ref{theorem:dmc} is achievable with arbitrarily low probability of error.

\section{Proof of Theorem \ref{theorem:gauss}}\label{sec:proof:gauss}
  In order to prove Theorem \ref{theorem:gauss}, we apply in this section Theorem \ref{theorem:dmc} to the complex Gaussian multiple relay network presented in Section
  \ref{sec:gaussian.channel}.
  \begin{itemize}
    \item \textbf{Rates on source messages:}
      Reconsider the source rate constraints given in (\ref{eq:sdd.mixed:theorem:110}) for the DMC. 
      Now it follows for $\I\left(\RV{V}_N^j; \RV{Y}_d | \RV{V}_N^{[1;j-1]}\right)$:
      \begin{align}
	\I\left(\RV{V}_N^j; \RV{Y}_d | \RV{V}_N^{[1;j-1]}\right) = &  \h\left(\RV{Y}_d | \RV{V}_N^{[1;j-1]}\right) - \h\left(\RV{Y}_d | \RV{V}_N^{[1;j]}\right), j\in[1;N]\\
	\begin{split}
	  = & \h\left(\sum\limits_{l=0}^{N} h_{l,d}\RV{X}_{l;b} + \RV{Z}_{d;b} - \sum\limits_{k=1}^{j-1}\left(h_{s,d}\sqrt{\Vfactor_{s,N}^k P_s} + 
	    \sum\limits_{l=k}^N h_{l,d}\sqrt{\Vfactor_{l,N}^k P_l} \right)\RV{V}_{N;b}^k\right) - \\
	  & \h\left(\sum\limits_{l=0}^{N} h_{l,d}\RV{X}_{l;b} + \RV{Z}_{d;b} - \sum\limits_{k=1}^{j}\left(h_{s,d}\sqrt{\Vfactor_{s,N}^k P_s} +
	    \sum\limits_{l=k}^N h_{l,d}\sqrt{\Vfactor_{l,N}^k P_l}\right)\RV{V}_{N;b}^k\right)
	\end{split}
      \end{align}
      with the differential entropy \cite[Eq. (9.34)]{Cover.Thomas.1991}
      \begin{equation}
	\h\left(\RV{X}=\left(\RV{X}_1, \dots, \RV{X}_n\right)\right)=\log\left(\left(2\pi e \right)^n \det \Matrix{K}_{\RV{X}}\right),
      \end{equation}
      and $\Matrix{K}_{\RV{X}}$ being the covariance matrix of the multivariate Gaussian r.v. $\RV{X}$.
      Now reconsider the definitions in (\ref{eq:gauss:140}), (\ref{eq:gauss:141}) and (\ref{eq:gauss:144}) for $\Vpower_{l,l'}^k$
      and $\Wpower_l^{k,k'}$, respectively. With these definitions it follows
      \begin{align}
	\I\left(\RV{V}_N^j; \RV{Y}_d | \RV{V}_N^{[1;j-1]}\right) & = \log\left(
	\frac{ \sum\limits_{l=1}^N\left(h_{l,d}^2\Wfactor_l^{[1;M_l]} P_l + \sum\limits_{k=1}^l \Vpower_{l, d}^k\right) + h_{s,d}^2 \Vfactor_{s,s}^{[1;N+1]}P_s + N_d 
	  - \sum\limits_{k=1}^{j-1} \Vpower_{N, d}^k }
	  { \sum\limits_{l=1}^N\left(h_{l,d}^2\Wfactor_l^{[1;M_l]} P_l + \sum\limits_{k=1}^l \Vpower_{l, d}^k\right) + h_{s,d}^2 \Vfactor_{s,s}^{[1;N+1]}P_s + N_d 
	  - \sum\limits_{k=1}^{j} \Vpower_{N, d}^k }
	\right)\\
	& = \C\left( \frac{\Vpower_{N, d}^j}
	{ \mathbf{\Vpower}_{l\in[1; N-1], d}^{[1; l]} + \mathbf{\Vpower}_{N,d}^{[j+1;N]} + \Wpower_d^{d,d} + \Gamma_{s,d}^{[1;N+1]} + N_d }
	\right)
      \end{align}
      with the capacity function $\C(x)=\log(1+x)$.
      In a similar way we can show
      \begin{multline}
	\I\left(\RV{V}_l^j; \RV{Y}_{l+k} | \RV{V}_l^{[1;j-1]},
	\left\{\RV{V}_i^{[1;i]}, \RV{W}_i^{[1;\inverseOrder_i(l+k)]} : i\in[l+1; l+k]\right\}, \RV{V}_{[l+k+1; N]}^{[1;l+k]}\right)\\
	< \C\left( 
	  \frac{\Vpower_{l, l+k}^j}
	  {\mathbf{\Vpower}_{l'\in[1;l-1],l+k}^{[1;l']} + \Vpower_{l,l+k}^{[j+1; l]} + \Vpower_{l'\in[l+k+1;N],l+k}^{[l+k+1;l']} + \Wpower_{l+k}^{l+1,l+k} + \Gamma_{s,l+k}^{[1;N+1]} + N_{l+k}}
	\right),\\
	l\in[1;N-1], j\in[1;l], k\in[1; N-l+1].
      \end{multline}
      
      Finally, reconsider the definitions in (\ref{eq:gauss:142}), (\ref{eq:gauss:143}) and (\ref{eq:gauss:145})-(\ref{eq:gauss:148}) which are used
      to derive the first term in (\ref{eq:sdd.mixed:theorem:110}):
      \begin{eqnarray}
	\lefteqn{\I\left(\RV{U}_{s}^j; \RV{Y}_k, \RV{\mathbf{\hat Y}}_{i\in[1;j-1]}^{\inverseOrder_i(k)} | \RV{U}_s^{[1;j-1]},
	\left\{\RV{V}_{[i;N]}^i, \RV{W}_i^{[1;\inverseOrder_i(k)]}: i\in[1;k]\right\}\right) = }\\ 
	  & & \h\left(\RV{Y}_k, \RV{\mathbf{\hat Y}}_{i\in[1;j-1]}^{\inverseOrder_i(k)} | \RV{U}_s^{[1;j-1]}, \left\{\RV{V}_{[i;N]}^i, \RV{W}_i^{[1;\inverseOrder_i(k)]}: i\in[1;k]\right\}\right) -\\
	  & & \h\left(\RV{Y}_k, \RV{\mathbf{\hat Y}}_{i\in[1;j-1]}^{\inverseOrder_i(k)} | \RV{U}_s^{[1;j]}, \left\{\RV{V}_{[i;N]}^i, \RV{W}_i^{[1;\inverseOrder_i(k)]}: i\in[1;k]\right\}\right)\\
	  & = & \log\left(\left(2\pi e\right)^j \det K_{s,k}^{j-1, j-1}\right) - \log\left(\left(2\pi e\right)^j \det K_{s,k}^{j-1, j}\right)\\
	  & = & \log\left(\frac{\det K_{s,k}^{j-1, j-1}}{\det K_{s,k}^{j-1, j}}\right).
      \end{eqnarray}
      Now it follows for the source rates $R_s^k$ using the previous results:
      \begin{multline}
	R_s^k < \min\limits_{l\in[k;N+1]} \log\left(\frac{\det K_{s,l}^{k-1, k-1}}{\det K_{s,l}^{k-1, k}}\right)\\
	+ \sum\limits_{j=k}^{l-1}\C\left(
	\frac{\Vpower_{j,l}^k}
	{ \mathbf{\Vpower}_{l'\in[1;j-1],l}^{[1;l']} + \Vpower_{j,l}^{[k+1; j]} + \Vpower_{l'\in[l+1;N],l}^{[l+1;l']} + \Wpower_{l}^{j+1,l} + \Gamma_{s,l}^{[1;N+1]} + N_{l} }
	\right).
      \end{multline}

    \item \textbf{Rates on broadcast messages:}
      Using the previous description it is immediately possible to state for the broadcast message constraints in (\ref{eq:sdd.mixed:120}):
      \begin{align}
	\begin{split}
	  \hat R_l^j < & \min\limits_{\begin{array}{l}k\in[j; M_l]\\ k'=o_l(k)\end{array}}\C\left(
	    \frac{h_{l,k'}^2\beta_l^jP_l}
	    {\mathbf{\Vpower}_{l'\in[1;l-1],k'}^{[1;l']} + \Vpower_{l'\in[k'+1;N],k'}^{[k'+1;l']} + \Wpower_{k'}^{l+1,k'}-h_{l,k'}^2\sum\limits_{i=1}^{j}\beta_l^iP_l + \Gamma_{s,k'}^{[1;N+1]} + N_{k'}}
	    \right),\\
	  & l\in[1;N+1], j\in[1;M_l].
	\end{split}\label{eq:sdd.mixed:gauss:130}
      \end{align}

    \item \textbf{Rates on successive refinement conditions:}
      Reconsider the necessary side condition on the successive refinement conditions given in (\ref{eq:sdd.mixed:130}).
      We can significantly simplify this condition to
      \begin{align}
	\begin{split}
	  \sum\limits_{i=1}^m \hat R_l^i > &
	  \max\limits_{\substack{j\in[m;M_l]\\ j'=o_l(j)}}
	    \I\Bigl(\RV{\hat Y}_l^m; \RV{Y}_l | \RV{Y}_{j'},
	    \RV{U}_s^{[1;l]}, \left\{\RV{V}_{[i;N]}^i, \RV{W}_i^{[1;\inverseOrder_i(j')]}: i\in[1;j']\right\}\Bigr), l\in[1; N], m\in[1; M_l],
	\end{split}\label{eq:sdd.mixed:gauss:150}
      \end{align}
      which is a result of the following inequality
      \begin{eqnarray}
	\I\left(\RV{\hat X}; \RV{Y}_l | \RV{Y}_{j'}, \RV{Y}, \RV{W}, \RV{U}\right) & \stackrel{(a)}{=} & \I\left(\RV{\hat X}; \RV{Y}_l | \RV{U}\right) - \I\left(\RV{\hat X}; \RV{Y}_{j'}, \RV{Y}, \RV{W} | \RV{U}\right)\\
	& \stackrel{(b)}{=} & \I\left(\RV{\hat X}; \RV{Y}_l | \RV{U}\right) - \I\left(\RV{\hat X}; \RV{Y}_{j'}, \RV{W} | \RV{U}\right) - \underbrace{\I\left(\RV{\hat X}; \RV{Y} | \RV{Y}_{j'}, \RV{W}, \RV{U}\right)}_{\geq 0}\\
	& \stackrel{(c)}{\leq} & \I\left(\RV{\hat X}; \RV{Y}_l | \RV{Y}_{j'}, \RV{W}, \RV{U}\right)
      \end{eqnarray}
      with the mappings
      \begin{align*}
	\RV{W} & \mapsto \left\{\RV{V}_{[i;N]}^i, \RV{W}_i^{[1;\inverseOrder_i(j')]}: i\in[l+1;j']\right\}, 
	\RV{\mathbf{W}}_{i\in[1;l]}^{[2;\inverseOrder_i(j')]}, & \RV{Y} & \mapsto \RV{\mathbf{\hat Y}}_{i\in[1;l-1]}^{\inverseOrder_i(j')},\\
	\RV{U} & \mapsto \left(\RV{U}_s^{[1;l]}, \left\{\RV{V}_{[i;N]}^i, \RV{W}_i^1: i\in[1;l]\right\}\right), & 
	\RV{X} & \mapsto \RV{\hat Y}_{l}^m.
      \end{align*}
      Furthermore, $(a)$ and $(c)$ follow the arguments already applied in (\ref{eq:sdd.mixed:160}) and $(b)$ uses the chain rule for mutual information.
      Let us again use the previous mappings and consider the following equalities:
      \begin{eqnarray}
	\I\left(\RV{\hat X}; \RV{Y}_l | \RV{Y}_{j'}, \RV{W}, \RV{U}\right) & = & \h\left(\RV{\hat{Y}}_l^m | \RV{Y}_{j'}, \RV{U}, \RV{W}\right) - \h\left(\RV{\hat{Y}}_l^m | \RV{Y}_l, \RV{Y}_{j'}, \RV{U}, \RV{W}\right)\label{eq:sdd.mixed:gauss:161}\\
	& \stackrel{(a)}{=} & \h\left(\RV{\hat{Y}}_l^m | \RV{Y}_{j'}, \RV{U}, \RV{W}\right) - \h\left(\RV{\hat{Y}}_l^m | \RV{Y}_l, \RV{U}\right)\label{eq:sdd.mixed:gauss:162}\\
	& = & \h\left(\RV{\hat{Y}}_l^m, \RV{Y}_{j'} | \RV{U, W}\right) - \h\left(\RV{Y}_{j'} | \RV{U, W}\right) - \h\left(\RV{\hat{Y}}_l^m | \RV{Y}_l, \RV{U}\right)\label{eq:sdd.mixed:gauss:163},
      \end{eqnarray}
      where $(a)$ exploits the Markov chain $\RV{Y}_l^m \Markov \left(\RV{Y}_l, \RV{U}\right)\Markov \left(\RV{Y}_{j'}, \RV{W}\right)$.
      The third term in (\ref{eq:sdd.mixed:gauss:163}) is given by
      \begin{eqnarray}
	\h\left(\RV{\hat{Y}}_l^m | \RV{Y}_l, \RV{U}\right) & = & \h\left(\RV{Y}_{l;b} + \sum\limits_{i=m}^{M_l}Z_{l;b}^i | \RV{Y}_{l;b}, \RV{U}\right)\\
	& = & \log\left(2\pi e \sum\limits_{i=m}^{M_l} N_l^i\right).\label{eq:sdd.mixed:gauss:170}
      \end{eqnarray}
      The second term in (\ref{eq:sdd.mixed:gauss:163}) is given in the same way by
      \begin{equation}
	\h\left(\RV{Y}_{j'} | \RV{U, W}\right) = \log\left(2\pi e\left(\Vpower_{l'\in[j'+1;N], j'}^{[j'+1;l']} + \Vpower_{s,j'}^{[l+1;N+1]} + 
	\Wpower_{j'}^{1,j'} + N_{j'}\right)\right).\label{eq:sdd.mixed:gauss:180}
      \end{equation}
      Finally, the first term in (\ref{eq:sdd.mixed:gauss:163}) is given by
      \begin{equation}
	\h\left(\RV{\hat{Y}}_l^m, \RV{Y}_{j'} | \RV{U, W}\right) = \log\left(\left(2\pi e\right)^2 \det \Matrix{K}_{l, j'}^m\right),\label{eq:sdd.mixed:gauss:190}
      \end{equation}
      with the covariance matrix $\Matrix{K}_{l,j'}^m$ given by
      \begin{align}
	\MatrixElement{1}{1}{\Matrix{K}_{l,j'}^m} = & \Var\left(\RV{\hat Y}_l^m | \RV{U, W}\right) = \Vpower_{l'\in[j'+1;N], l}^{[j'+1;l']} + \Vpower_{s,l}^{[l+1;N+1]} + 
	  \Wpower_{l}^{1,j'} + N_{l} + \sum\limits_{i=m}^{M_l} N_l^i \\
	\MatrixElement{2}{2}{\Matrix{K}_{l,j'}^m} = & \Var\left(\RV{Y}_{j'} | \RV{U, W}\right) = \Vpower_{l'\in[j'+1;N], j'}^{[j'+1;l']} + \Vpower_{s,j'}^{[l+1;N+1]} + 
	  \Wpower_{j'}^{1,j'} + N_{j'}\\
	\MatrixElement{1}{2}{\Matrix{K}_{l,j'}^m} = & \Cov\left(\RV{\hat Y}_l^m | \RV{U, W}; \RV{Y}_{j'} | \RV{U, W}\right) = \VpowerCov_{l'\in[j'+1;N], j', l}^{[j'+1;l']} + \VpowerCov_{s,j',l}^{[l+1;N+1]},
      \end{align}
      where we used the definitions given in (\ref{eq:gauss:142}) and (\ref{eq:gauss:143}) for $\VpowerCov_{l,m,m'}^k$.
      Finally, using (\ref{eq:sdd.mixed:gauss:170})-(\ref{eq:sdd.mixed:gauss:190}) in (\ref{eq:sdd.mixed:gauss:150}) it follows that
      \begin{equation}
	\sum\limits_{i=1}^m \hat R_l^i > \max\limits_{\substack{j\in[m;M_l]\\ j'=o_l(j)}}\log\left(\frac{\det K_{l,j'}^m}{\left(\sum\limits_{i=m}^{M_l} N_l^i\right)\cdot
	  \left(\Vpower_{l'\in[j'+1;N], j'}^{[j'+1;l']} + \Vpower_{s,j'}^{[l+1;N+1]} + \Wpower_{j'}^{1,j'} + N_{j'}\right)}\right)\label{eq:sdd.mixed:gauss:210}.
      \end{equation}
      Since determinant of $\Matrix{K}_{l,j'}^m$ is given by
      \begin{equation}
	\det K_{l,j'}^m = \Var\left(\RV{\hat Y}_l^m | \RV{U, W}\right)\Var\left(\RV{Y}_{j'} | \RV{U, W}\right) - \Cov\left(\RV{\hat Y}_l^m | \RV{U, W}; \RV{Y}_{j'} | \RV{U, W}\right)^2,
      \end{equation}
      it follows for (\ref{eq:sdd.mixed:gauss:210}):
      \begin{multline}
	\sum\limits_{i=1}^m \hat R_l^i > \max\limits_{\substack{j\in[m;M_l]\\ j'=o_l(j)}} = \C\left(
	  \frac{\Vpower_{l'\in[j'+1;N], l}^{[j'+1;l']} + \Vpower_{s,l}^{[l+1;N+1]} + \Wpower_l^{1,j'} + N_{l}}{\sum\limits_{i=m}^{M_l} N_l^i} -\right.\\
	  \left.\frac{\left(\VpowerCov_{l'\in[j'+1;N], j', l}^{[j'+1;l']} + \VpowerCov_{s,j',l}^{[l+1;N+1]}\right)^2}{\left(\sum\limits_{i=m}^{M_l} N_l^i\right)\cdot\left(\Vpower_{l'\in[j'+1;N], j'}^{[j'+1;l']} + \Vpower_{s,j'}^{[l+1;N+1]} + \Wpower_{j'}^{1,j'} + N_{j'}\right)}
	  \right)
	  \label{eq:sdd.mixed:gauss:220}
      \end{multline}
      Using (\ref{eq:sdd.mixed:gauss:220}) we get the constraints on the quantization noise variance given in (\ref{eq:gauss:170}).
  \end{itemize}

\end{document}